\documentclass[11pt,]{article}
\usepackage{lmodern}
\usepackage{natbib}
\usepackage[a4paper, total={7in, 9in}]{geometry}
\usepackage{wrapfig}
\usepackage{amsmath,amsthm,amssymb}

\AtBeginDocument{}
\usepackage{sectsty}
\allsectionsfont{\centering}
\usepackage{titlesec}
\usepackage{booktabs}
\titlelabel{\thetitle.\quad}
\usepackage{microtype} 
\usepackage{filecontents}
\usepackage{enumitem}   
\usepackage{adjustbox}
\usepackage{xurl}
\usepackage{subcaption}
\usepackage{color, colortbl}
\usepackage{geometry}
\providecommand{\keywords}[1]{\textbf{Key words:} #1}
\definecolor{Gray}{gray}{0.9}
\usepackage{subfiles}

\usepackage{titling}
\usepackage{authblk}

\title{A predictive model for planning emergency events rescue during COVID-19 in Lombardy, Italy}

\author[1]{Angela Andreella}
\affil[1]{Universit\`a Ca' Foscari Venezia}
\author[2]{Antonietta Mira}
\affil[2]{Universit\`a della Svizzera Italiana}
\author[3]{Spyros Balafas}
\affil[3]{Universit\`a degli studi dell'Insubria}
\author[2]{Ernst C. Wit}
\author[4]{Fabrizio Ruggeri}
\affil[4]{Consiglio Nazionale delle Ricerche, Istituto di Matematica Applicata e Tecnologie Informatiche}
\author[5]{Giovanni Nattino}
\author[5]{Giulia Ghilardi}
\author[5]{Guido Bertolini}
\affil[5]{Istituto di Ricerche Farmacologiche Mario Negri, IRCCS}

\date{}
\usepackage{array}
\usepackage{multirow}
\usepackage{float}
\usepackage{tabu}
\usepackage{threeparttable}
\usepackage{threeparttablex}
\usepackage[normalem]{ulem}
\usepackage{makecell}
\usepackage{xcolor}
\usepackage{algorithm2e}
\usepackage[T1]{fontenc}
\usepackage[utf8]{inputenc}
\usepackage{array}
\usepackage{caption}
\usepackage{wrapfig}
\usepackage{colortbl}
\usepackage{pdflscape}
\usepackage{bibentry}

\begin{document}
\maketitle

\begin{abstract}
Italy, particularly the Lombardy region, was among the first countries outside of Asia to report cases of COVID-19. The emergency medical service called Regional Emergency Agency (AREU) coordinates the intra- and inter-regional non-hospital emergency network and the European emergency number service in Lombardy. AREU must deal with daily and seasonal variations of call volume. The number and type of emergency calls changed dramatically during the COVID-19 pandemic. A model to predict incoming calls and how many of these turn into events, i.e., dispatch of transport and equipment until the rescue is completed, was developed to address the emergency period. We used the generalized additive model with a negative binomial family to predict the number of events one, two, five, and seven days ahead. The over-dispersion of the data was tackled by using the negative binomial family and the nonlinear relationship between the number of events and covariates (e.g., seasonal effects) by smoothing splines. The model coefficients show the effect of variables, e.g., the day of the week, on the number of events and how these effects change during the pre-COVID-19 period. The proposed model returns reasonable mean absolute errors for most of the 2020-2021 period.
	
	\keywords{Covid-19, emergency departments data, emergency call data, predictive models, decision support systems}
\end{abstract}

\section{Introduction}
\begin{figure}
	\centering
	\begin{subfigure}[b]{0.4\textwidth}
		\includegraphics[width=\textwidth]{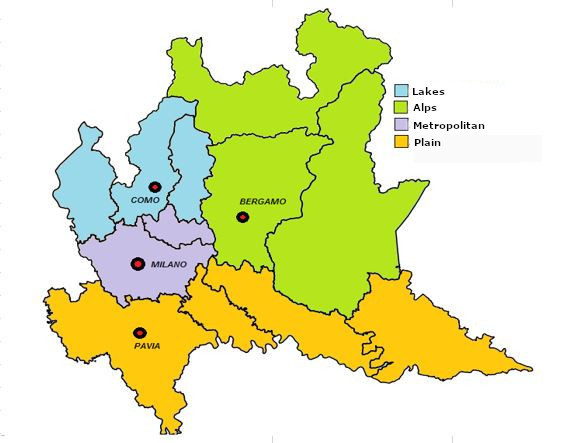}\\(a)
	\end{subfigure}
	\begin{subfigure}[b]{0.4\textwidth}
		\includegraphics[width=\textwidth]{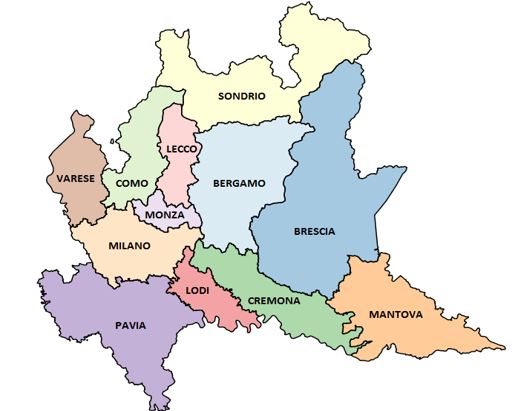}\\(b)
	\end{subfigure}
\caption{Description of the AREU organizational structures: (a) $4$ SOREUs: AREU emergency management; (b) $12$ ATTs: AREU management of territorial resources. 	\label{fig0}}
\end{figure}

The emergency medical service (EMS) of the Lombardy region of Italy is coordinated by the Regional Emergency Agency (AREU). It guarantees the intra- and inter-regional coordination, guidance, management, performance, and monitoring of the non-hospital emergency network and the NUE (European emergency number) service, for which it is responsible for planning and control. AREU is organized into peripheral structures called Territorial Business Units (AAT) and Regional Emergency Operations Rooms (SOREU). The twelve AATs are distributed over the Lombardy region approximately following the provincial division, while the four SOREUs coordinate rescue operations in supra-provincial competence areas (i.e., AAT aggregations). The geographical position and organization of the SOREUs and the distribution of the AATs are represented in Figure \ref{fig0}. 

AREU must deal with daily and seasonal variations in call volume. In addition, many factors can describe the call volume across time beyond the annual trend, such as social and demographic factors \citep{kamenetzky1982estimating}, abnormal weather conditions \citep{alessandrini2011emergency} like heat waves or cyclones, and epidemiological factors \citep{diaz2001model} as flu incidence. Factors related to the day of the week, time of the day, seasonal and yearly variations that characterize the pattern of time series \citep{cantwell2015time, batal2001predicting} must also be taken into account.

Furthermore, the number and type of calls to the emergency call center changed dramatically during the COVID-19 pandemic. Therefore, the development of statistical modeling becomes essential for AREU to predict the number of incoming calls and how many of these turn into an actual event -- dispatch of the most appropriate transport until the rescue is completed and/or the patient is referred to the most suitable hospital facilities--. The development of a valuable predictive model is crucial in this emergency period to have an accurate organization of the actions towards the solution of an emergency event. The change in the daily volume of emergency calls impedes the rationalization of resource allocation, and this variation has increased due to the COVID-19 pandemic. 

The paper is organized as follows. In Section~\ref{data} we describe the emergency response data provided by the AREU. In Section~\ref{model} we describe the model, focusing on the \emph{Plain} SOREU. Section~\ref{results} outlines the predictive performance of the model for each SOREU.

\section{Methods}\label{method}

\subsection{Data}\label{data}
The data used are composed of a plethora of variables describing the Lombardy emergency events from January $1$st, $2015$ until May $9$th, $2021$. In particular, we employ $134$ variables which are gathered from different sources: AREU for emergency response data, Regional Agency for Environmental Protection (ARPA) for weather data, Department of Civil Protection for COVID-19 data, and Higher Institute of Health (ISS) for epidemiological information.

The data from AREU consists of information about all calls received: the SOREU receiving the call, the exact time of the call (date and hour/minutes), the call classification (first aid, etc.), the AREU administrative area (province, zone, AAT), 
the location where the call was initiated (home, street, etc.) and its geographic coordinates, the reason of the call, the severity code (triage), and if the call activated an aid response, i.e., it became an event. The data are aggregated at an hourly level. 

The ARPA data are available through the Open Data project \citep{openData}, which provides weather data (temperature, rainfall, and snowfall) collected from sensors located across the Lombardy region. ARPA relies on a vast network of monitoring stations throughout Lombardy. The most central sensors in the city center for each AAT are chosen, and the weighted average is computed to obtain the variable at the level of SOREU. The weights equal the proportion of calls (from January $1$st, $2015$ until May $9$th, $2021$) for each province in SOREU. 
The data are collected on an hourly basis, but are aggregated at daily level, computing their average.

The COVID-19 data from the Department of Civil Protection are available at \cite{dipCovid}. The data are at the level of the Lombardy region, except for the total number of positive cases at any given date, which is at the province level. With reference to the temporal level, data are available on a daily basis. In addition to the basic epidemiological information, we compute the effective reproduction number \citep{rodpothong2012viral,dabbaghian2014theories} based on the method used by the ISS, available at \cite{ISS}.

Finally, the flu weekly incidence at the national level from $2015$ to $2021$ is considered as communicated by ISS \citep{ISS}. 

\subsection{Model}\label{model}

Let $Y$ be a random vector of dimension $T\times1$ containing the observations $y_{t}$ representing the counts of events at time $t\in\mathcal{T}$, $\mathcal{T}:= \left\{1,\dots, T\right\}$, where $t$ is an hourly interval. Let $\mathbf{X}=\left(X_1, \dots, X_K \right)$ be a set of $T$-dimensional covariates associated with the response variable $Y$ that are indexed by $k=1, \dots,K$, which can be year, month or day specific, etc. Possible covariates can be lags of $Y$ such as the total number of events in the previous day. 

Since we are dealing with count data and we expect over-dispersion, we use the generalized additive model (GAM) \citep{SW1, SW2, SW3} with negative binomial family \citep{casella2021statistical}. In particular, the GAM model allows to capture the non linear relationship between $\mathbf{X}$ and $Y$ specifying smoothing splines.

Consider a negative binomial $Y_t\sim \mathcal{NB}\left(r_t, \pi\right)$, where $t = 1, \dots, T$, with $\text{E}\left(Y_t\right) = \mu_t=\frac{r_t \pi}{1-\pi}$ and $\text{V}\left(Y_t\right) = \mu_t + \frac{\mu_t^2}{\pi}$. The negative binomial GAM model $\mu_t $ is described as follows:
\begin{equation*}
\ln(\mu_t) = \alpha + \sum_{k = 1}^{K} f_k(X_{tk}) \quad \text{where} \quad f_{k}(X_{tk}) = \sum_{d = 1}^{D_k} \beta_{kd} b_{kd}(X_{tk}).
\end{equation*}
where $\beta_{kd}$ are the unknown parameters to be estimated, $b_{kd}(\cdot)$ are known basis functions, and $D_{k}$ is the number of basis for the covariate $X_{tk}$.

To find the optimal model to predict the number of events, we cross-validate five models having different sets of covariates $\mathbf{X}$ across different time periods, focusing on the \emph{Plain} SOREU. We choose this SOREU being the second largest area after the \emph{Metropolitan} SOREU for which AREU already had a predictive model. We use the mean absolute error (MAE) computed across $N$ days as performance metric:
\begin{equation*}
    MAE = \dfrac{\sum_{i = 1}^{N}|E_i|}{N}
\end{equation*}
where $|\cdot|$ is the absolute value, $E_i = \frac{\hat{Y_i} - Y_i}{Y_i}$ is the prediction error of interest for AREU, $Y_i$ and $\hat{Y_i}$ are the observed and predicted values for day $i$. 

The optimal model includes the covariates described in Table \ref{table_cov}. Seasonal effects of hours and quarters are modelled by two cubic regression splines \citep{green2019nonparametric} with $24$ and $4$ basis functions, respectively. The P-spline \citep{eilers1996flexible} with $7$ basis functions is imposed for the day variable, and finally, the tensor product smooth \citep{ramsay1997spline} (two-dimensional smooth, where the shape of one dimension varies smoothly over the other dimension) between day and hour covariates is applied to analyze the interaction between these two variables. The remaining covariates enter linearly in the model.

\begin{table}[]
\begin{tabular}{@{}ll@{}}
\textbf{Name}      & \textbf{Description} \\
\midrule
Hour                & Numeric variable describing the hour \\&of the day when the event occurs at SOREU level.                                                                                                           \\
\rowcolor{Gray}  

Day                & Numeric variable describing the day \\
\rowcolor{Gray} 
&of the week when the event occurs at SOREU level.
                         \\

Quarter                & Numeric variable describing the quarter \\& of the year when the event occurs at SOREU level. \\
\rowcolor{Gray}  

Temperature      & Numeric variable describing the daily average \\
\rowcolor{Gray}  
&temperature at SOREU level.\\

events.lag1/events.lag2/events.lag3        & Numeric variable describing the number of events \\

& of the day before delayed respectively  $1$/$2$/$3$ hours. \\
& at SOREU level.                                                                                                                                                                                                                       \\
		\rowcolor{Gray}  

events.lagday1/events.lagday2/events.lagday7 & Numeric variable describing the number of events\\
		\rowcolor{Gray}  

&  aggregated by day and delayed respectively  $1$/$2$/$7$ days \\      
		\rowcolor{Gray}  
& at SOREU level.                                            \\

Rt     & Numeric variable describing the reproduction based on \\

& the number of the totalamount of positive cases lagged \\ &  one day at Lombardy level.\\
\rowcolor{Gray}  

Flu           & Numeric variable describing the flu incidence at weekly\\
\rowcolor{Gray}  

&and national level lagged one day.                                                         
\end{tabular}
\caption{\label{table_cov} Description of the covariates entering in the final model.}
\end{table}

For each covariate, we analyze the estimated association with the outcome. Figure \ref{fig1} illustrates the seasonal effects of the hours within a day. We notice that in the late morning and in the afternoon (when people are typically awake) the effect is positive, while during the night and in the early morning (when people are usually sleeping) it becomes negative. Figure \ref{fig11} shows the effects of the day of the week. Mondays seem to have the highest incidence of events compared to the rest of the week. Figure \ref{fig2} shows the seasonal effects of the quarters. The number of events is generally lower in the summer, when the schools are closed and people are on vacation. It is also probably associated with rising temperatures. Finally, looking at the interaction between hours and days in Figure \ref{fig21}, we can see a positive effect in the morning (9 a.m. - 12 p.m.) of weekdays, and a strong negative effect in the morning of the weekend. Furthermore, we can see a positive effect during the night in the weekend, and a negative effect around 5 a.m. during Monday, Tuesday, and Wednesday. 

\begin{figure}[tb]
	\centering
	\begin{subfigure}[b]{0.4\textwidth}
		\includegraphics[width=\textwidth]{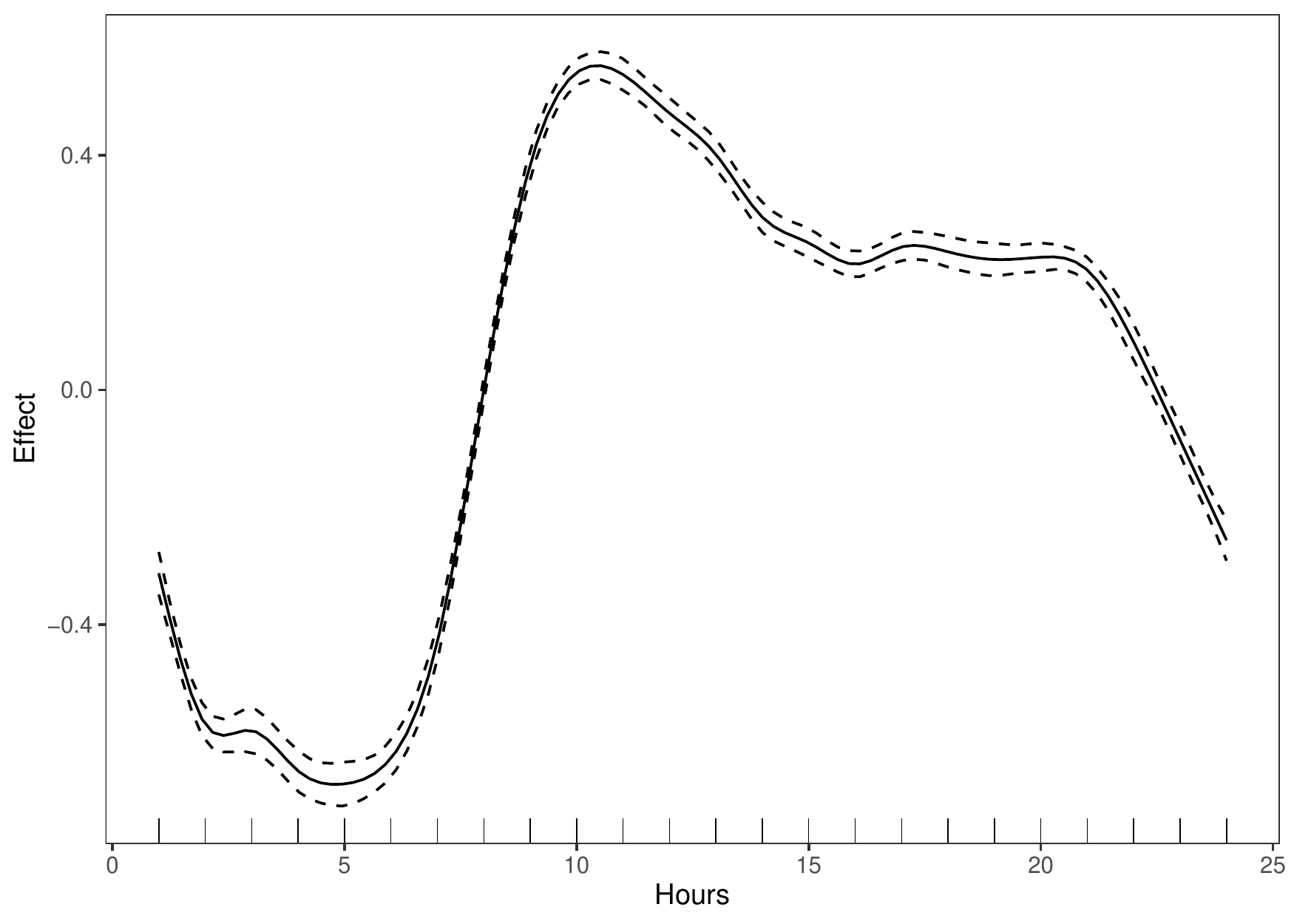}
		\caption{}
		\label{fig1}
	\end{subfigure}
	\begin{subfigure}[b]{0.4\textwidth}
		\includegraphics[width=\textwidth]{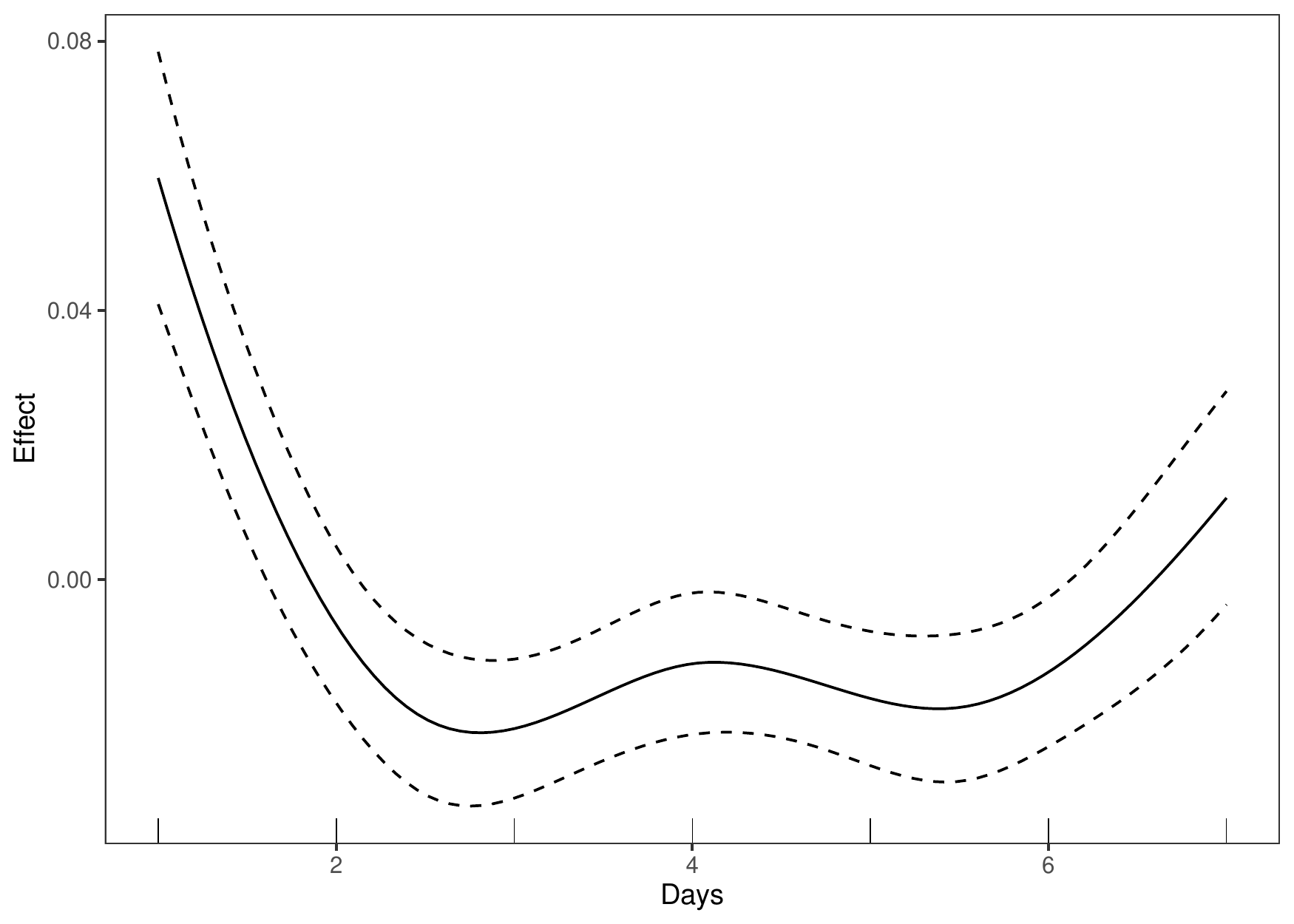}
		\caption{}
		\label{fig11}
	\end{subfigure}
	\begin{subfigure}[b]{0.4\textwidth}
		\includegraphics[width=\textwidth]{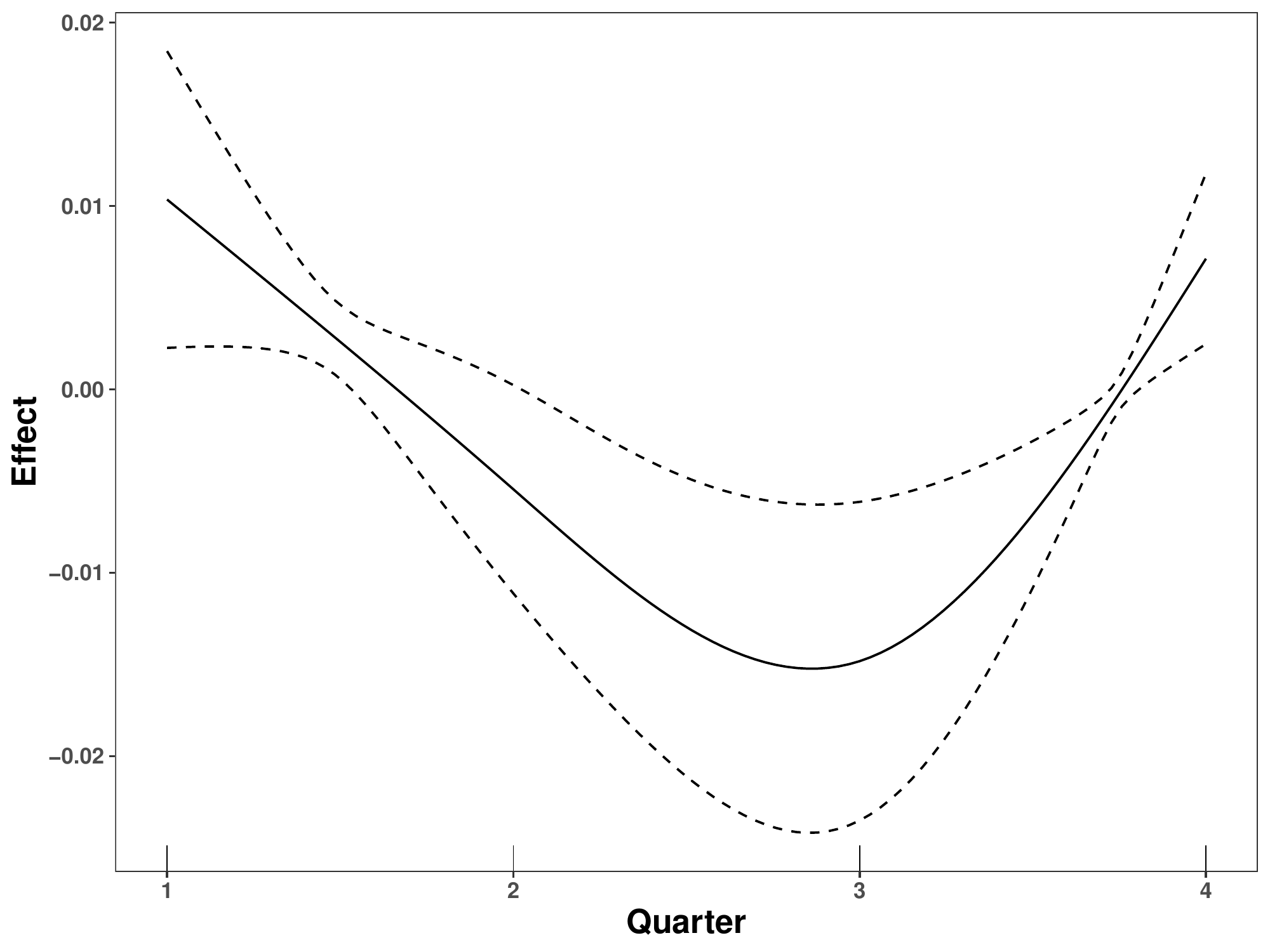}
		\caption{}
		\label{fig2}
	\end{subfigure}
	\begin{subfigure}[b]{0.4\textwidth}
		\includegraphics[width=\textwidth]{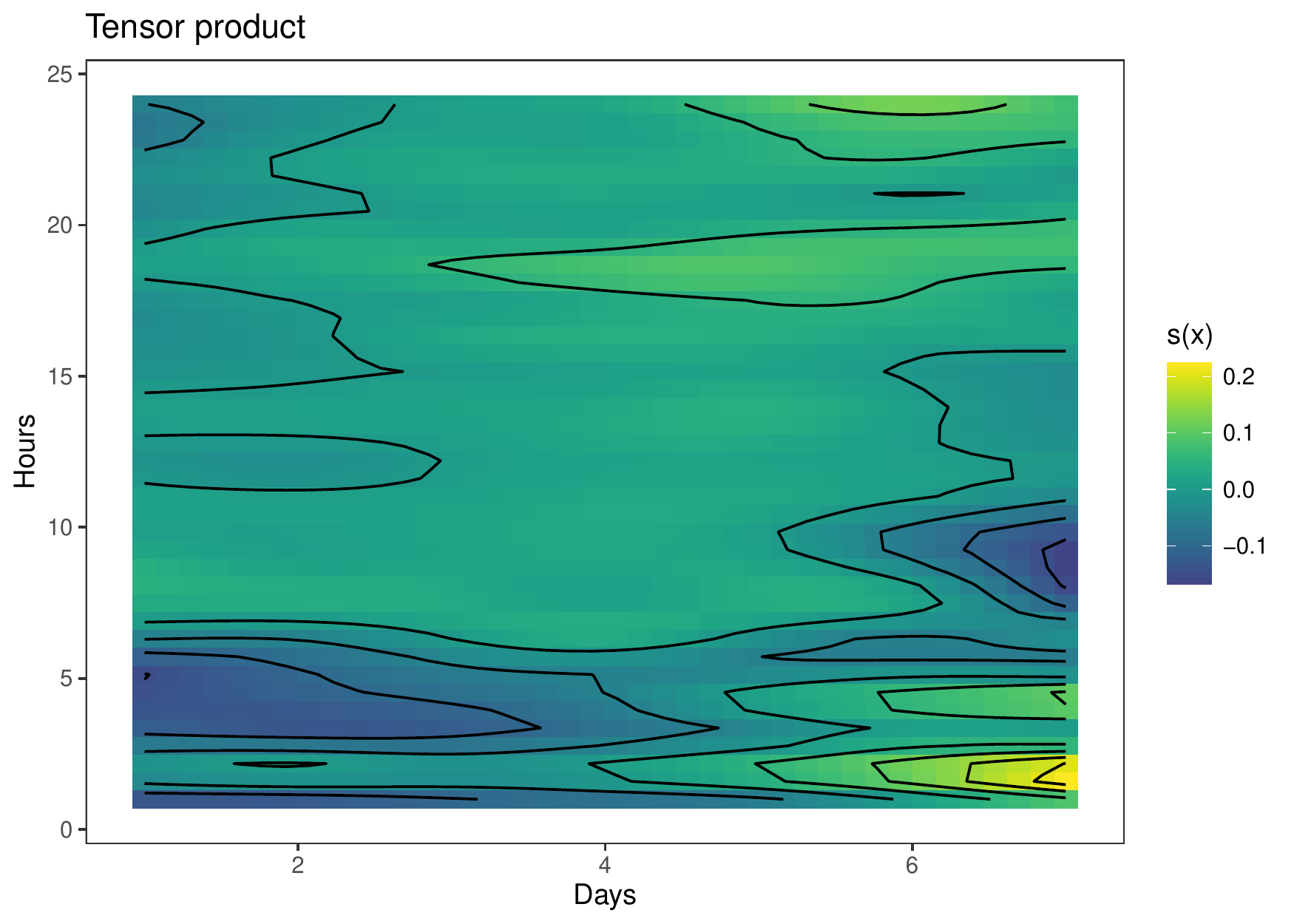}
		\caption{}
		\label{fig21}
	\end{subfigure}
\caption{\emph{Plain} model coefficients plots during the COVID-19 pandemic: (a) Effect of the hour of the day; (b) Effect of the day of the week, where $1$ stands for Monday; (c) Effect of the quarters of the year, where $1$ stands for the first quarter of the year; (d) Effect of the interaction between days and hours, where $1$ stands for Monday.}
\label{fig:coef}
\end{figure}

\begin{figure}[tb]
	\centering
	\begin{subfigure}[b]{0.4\textwidth}
		\includegraphics[width=\textwidth]{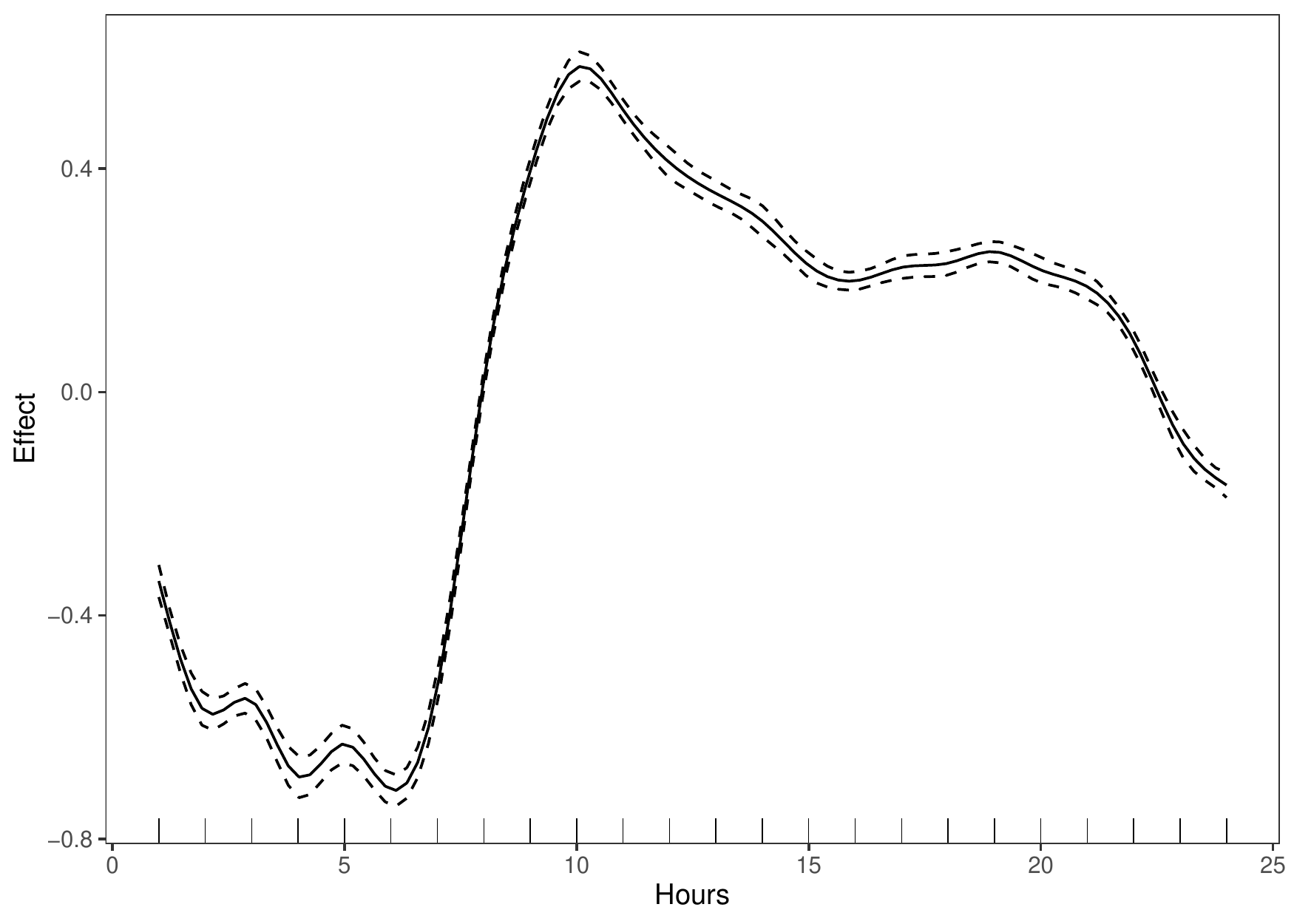}
		\caption{}
		\label{fig1pre}
	\end{subfigure}
	\begin{subfigure}[b]{0.4\textwidth}
		\includegraphics[width=\textwidth]{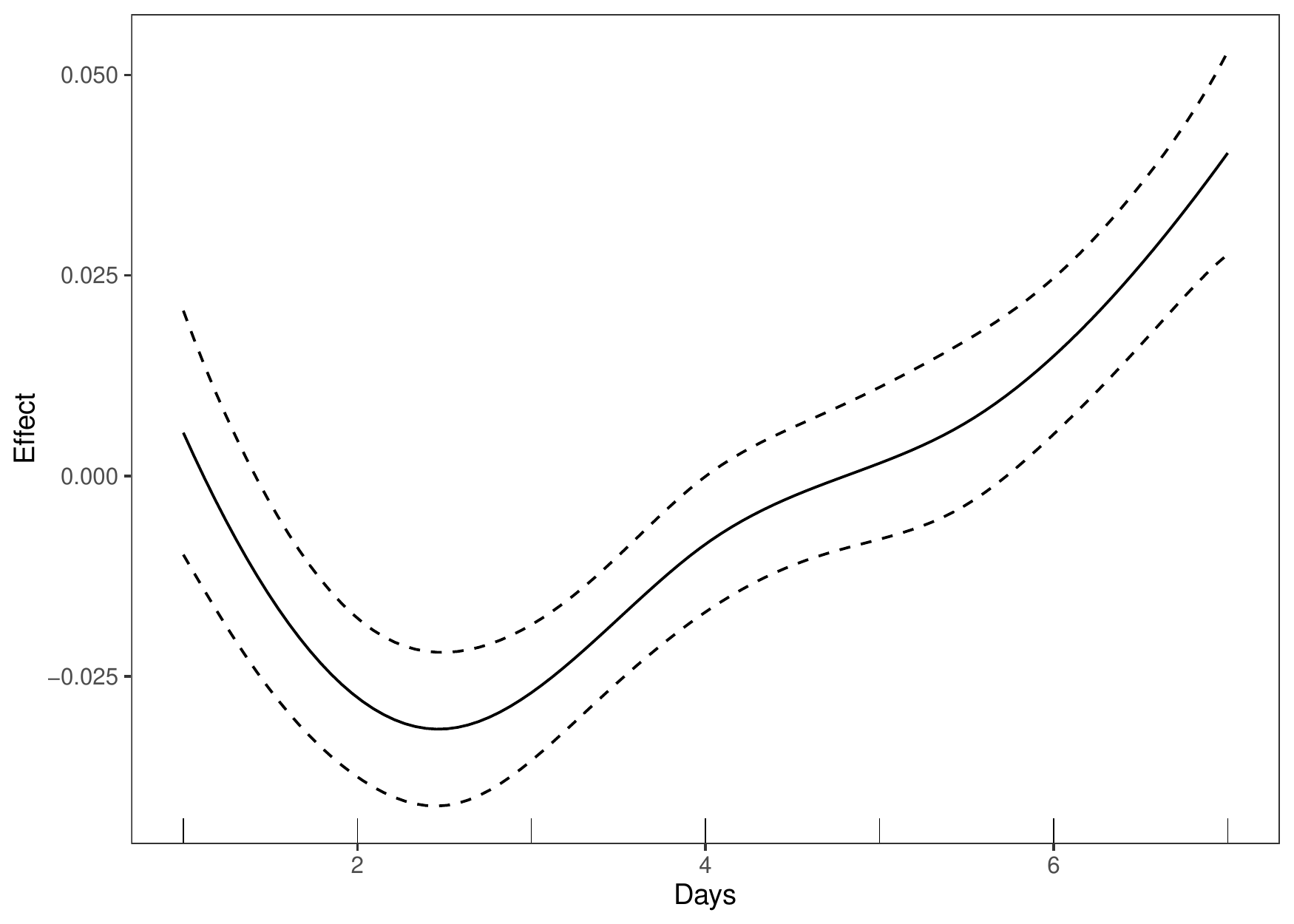}
		\caption{}
		\label{fig11pre}
	\end{subfigure}
	\begin{subfigure}[b]{0.4\textwidth}
		\includegraphics[width=\textwidth]{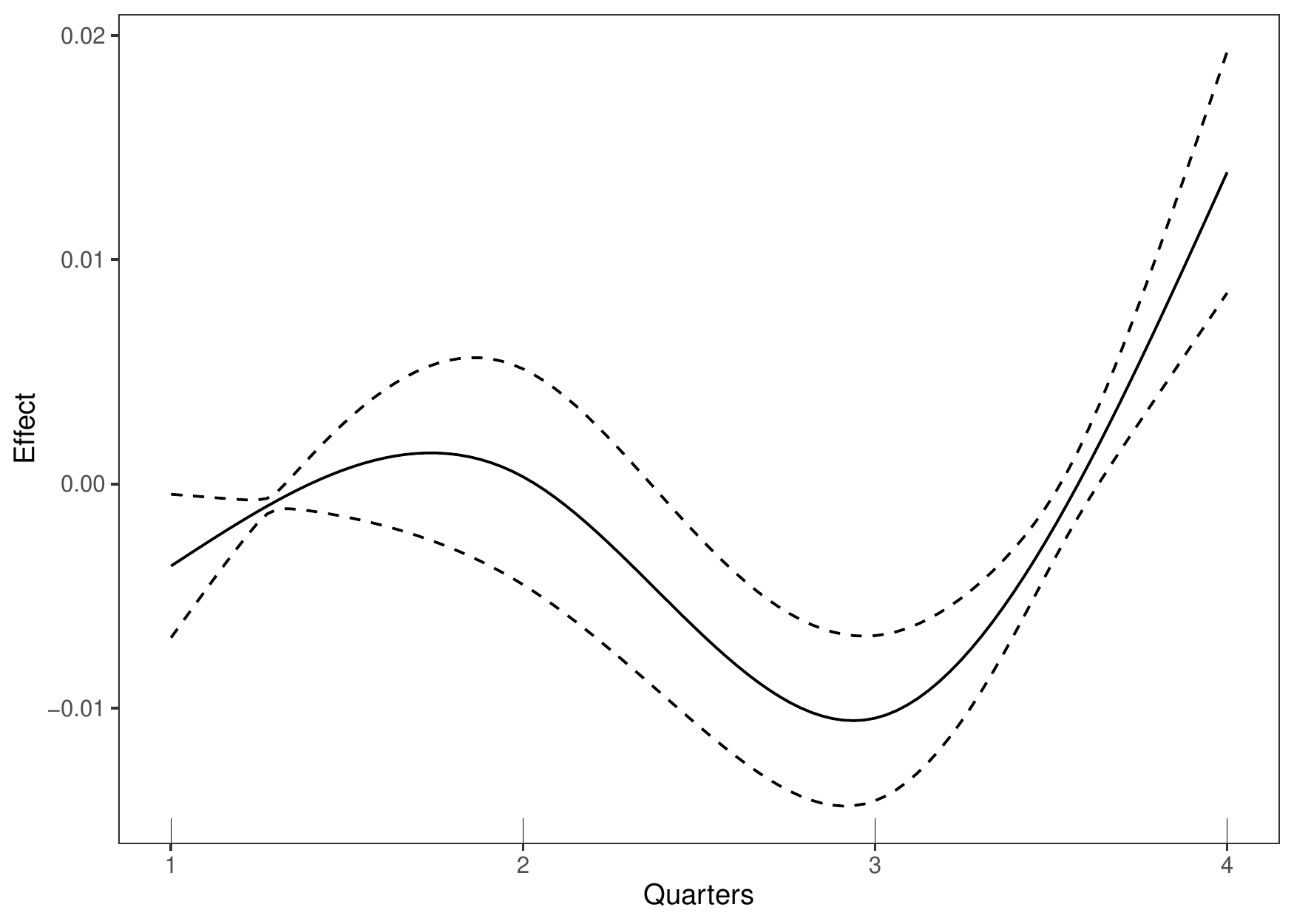}
		\caption{}
		\label{fig2pre}
	\end{subfigure}
	\begin{subfigure}[b]{0.4\textwidth}
		\includegraphics[width=\textwidth]{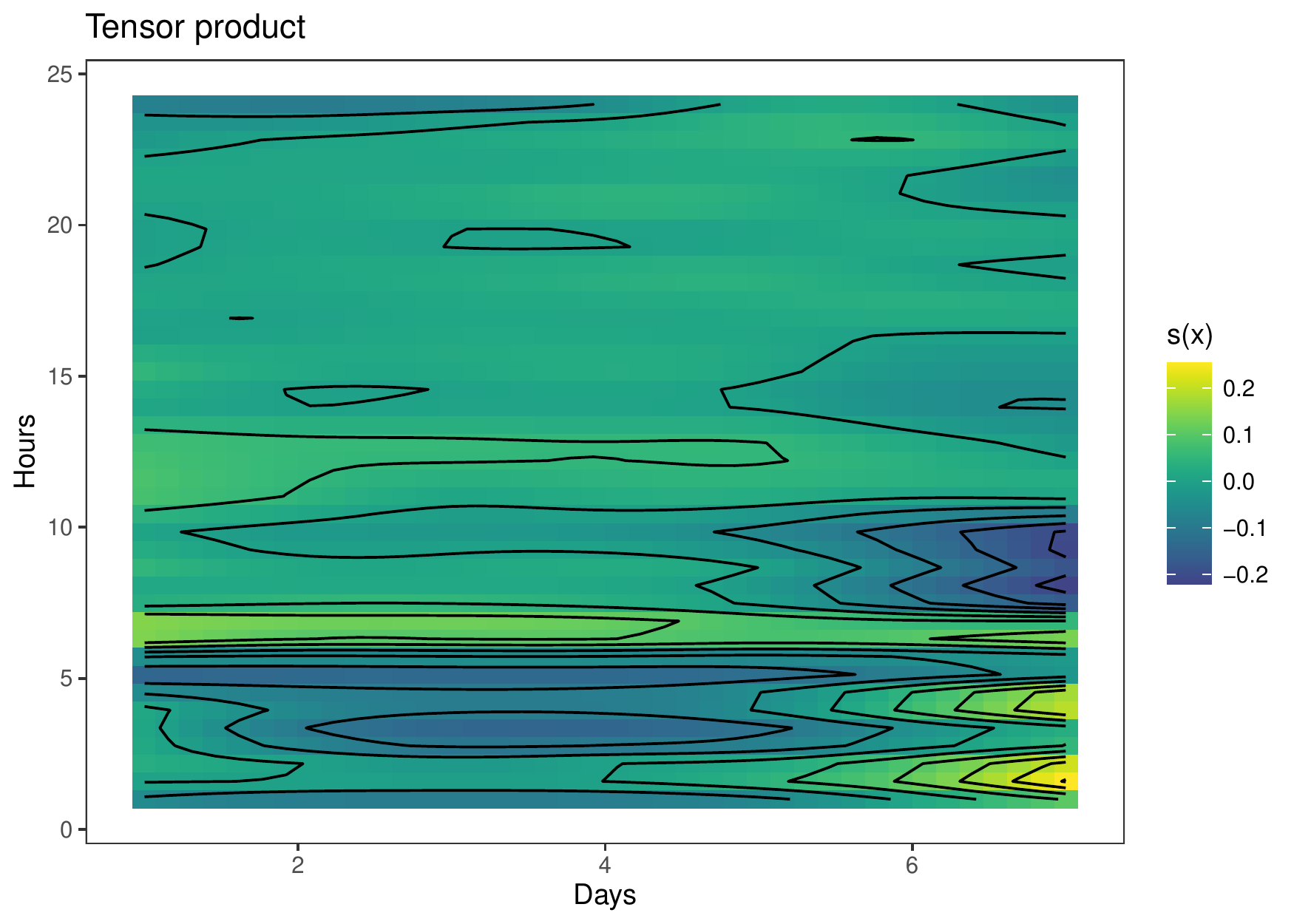}
		\caption{}
		\label{fig21pre}
	\end{subfigure}
\caption{\emph{Plain} model coefficients plots using data before COVID-19 pandemic: (a) Effect of the hour of the day; (b) Effect of the day of the week, where $1$ stands for Monday; (c) Effect of the quarters of the year, where $1$ indicates the first quarter of the year; (d) Effect of the interaction between days and hours, where $1$ stands for Monday.}
\label{fig:coef_preCovid}
\end{figure}

Finally, we estimate the model considering data before the COVID-19 era to see if the temporal dynamics, in particular the seasonal effects, have also changed due to the pandemic. Figure \ref{fig:coef_preCovid} shows the same plots presented in Figure \ref{fig:coef}. We can see that the effect of the hour of the day (Figure \ref{fig1pre}) and of the quarter (Figure \ref{fig21pre}) do not change markedly, unlike the effect of the day of the week (Figure \ref{fig11pre}). One possible interpretation could be that before the COVID-19 pandemic, many emergency events were concentrated during the weekend. Therefore, with the arrival of the COVID-19 pandemic, the emergencies referred to AREU have shifted by a few days. People become infected during the week and weekend and seek care on Monday.  Before COVID-19 pandemic, on the other hand, the primary demand, as we have said, was at weekends, when people went out to parties/hiking or when general practitioners were simply not available. Furthermore, Figure \ref{fig21pre} shows a positive effect on early mornings during all weeks, and a night effect during the weekend. Again, before COVID-19 pandemic, people most needed AREU interventions during weeknights and weekends. 

We use the open-source statistical software \texttt{R} \citep{Rcit}. From the available \texttt{R} packages for fitting GAM(s), we used the \texttt{mgcv} \citep{SW3} which permits to fit GAMs on large datasets thanks to the \texttt{bam} function. The full code used in this paper is available on \url{https://github.com/angeella/Tsunami_project}.

\section{Results}\label{results}

\begin{figure}[tb]
	\centering
		\begin{subfigure}[b]{0.4\textwidth}
	\includegraphics[width=\textwidth]{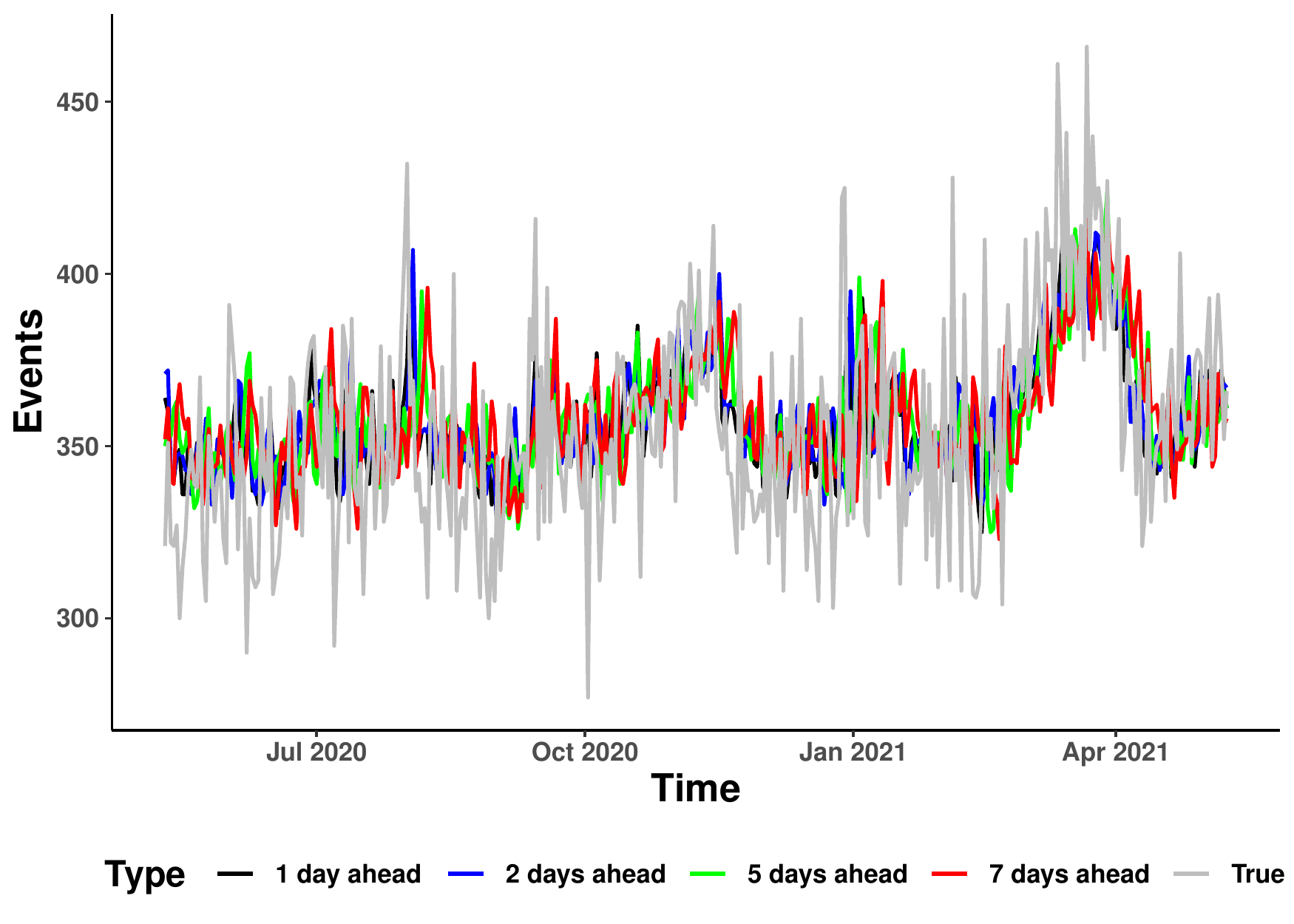}
	\caption{\label{fig:plot_totP} }
\end{subfigure}
	\centering
		\begin{subfigure}[b]{0.45\textwidth}
	\includegraphics[width=\textwidth]{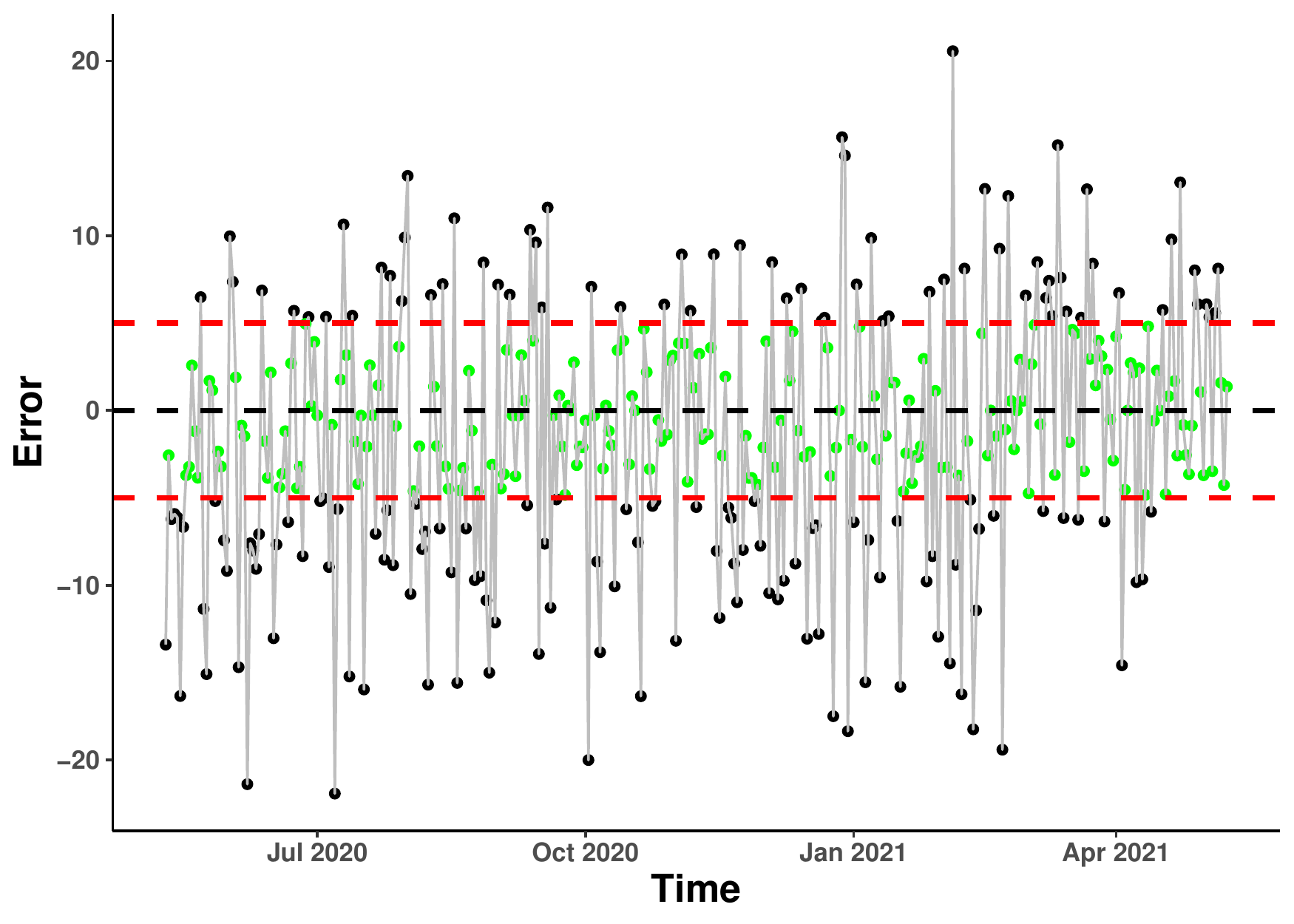}
	\caption{\label{fig:mse_1day}}
	\end{subfigure}
	\caption{\emph{Plain} model results: (a) Predictions one, two, five and seven days ahead across one year and relative true values; (b) Forecast errors $1$ day ahead, the predictions inside the two black dotted horizontal lines have absolute error below $5$. 	\label{figP}}
\end{figure}

First of all, we focus on reporting the results considering the \emph{Plain} SOREU being the one used to train the model.  
We fit the model presented in the previous section across one year, from May, $9$th, $2020$  until May, $9$th, $2021$, and each time we forecast the number of events one, two, five and seven days ahead. Figure \ref{fig:plot_totP} shows the predictions one, two, five, and seven days ahead obtained from the selected model with the true number of events, while Figure \ref{fig:mse_1day} shows the forecast errors across one year predicting one day ahead. The periods where the error appears to be slightly larger are during the second pandemic wave, November-February $2021$ around the holiday season. In general, the mean absolute error considering the whole year equals $4.53\%$. This is a good result considering that AREU required the mean absolute error to be at most $5\%$ for the predictions to be useful for efficiently planning their activities from a socio-economical point of view.

\begin{figure}[tb]
	\centering
			\begin{subfigure}[b]{0.4\textwidth}
	\includegraphics[width=\textwidth]{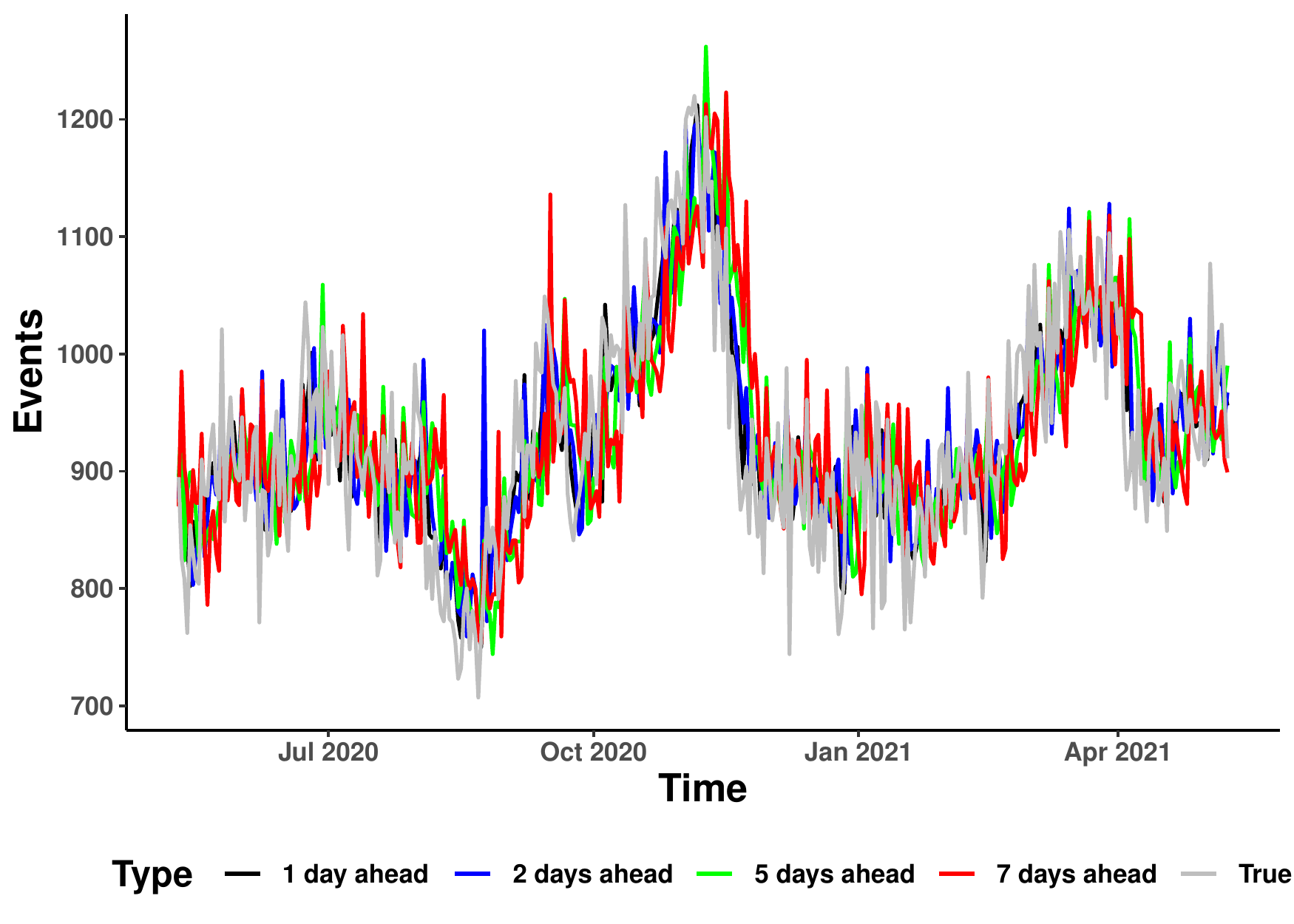}
	\caption{\label{fig:plot_totM} }
	\end{subfigure}
\begin{subfigure}[b]{0.45\textwidth}
	\includegraphics[width=\textwidth]{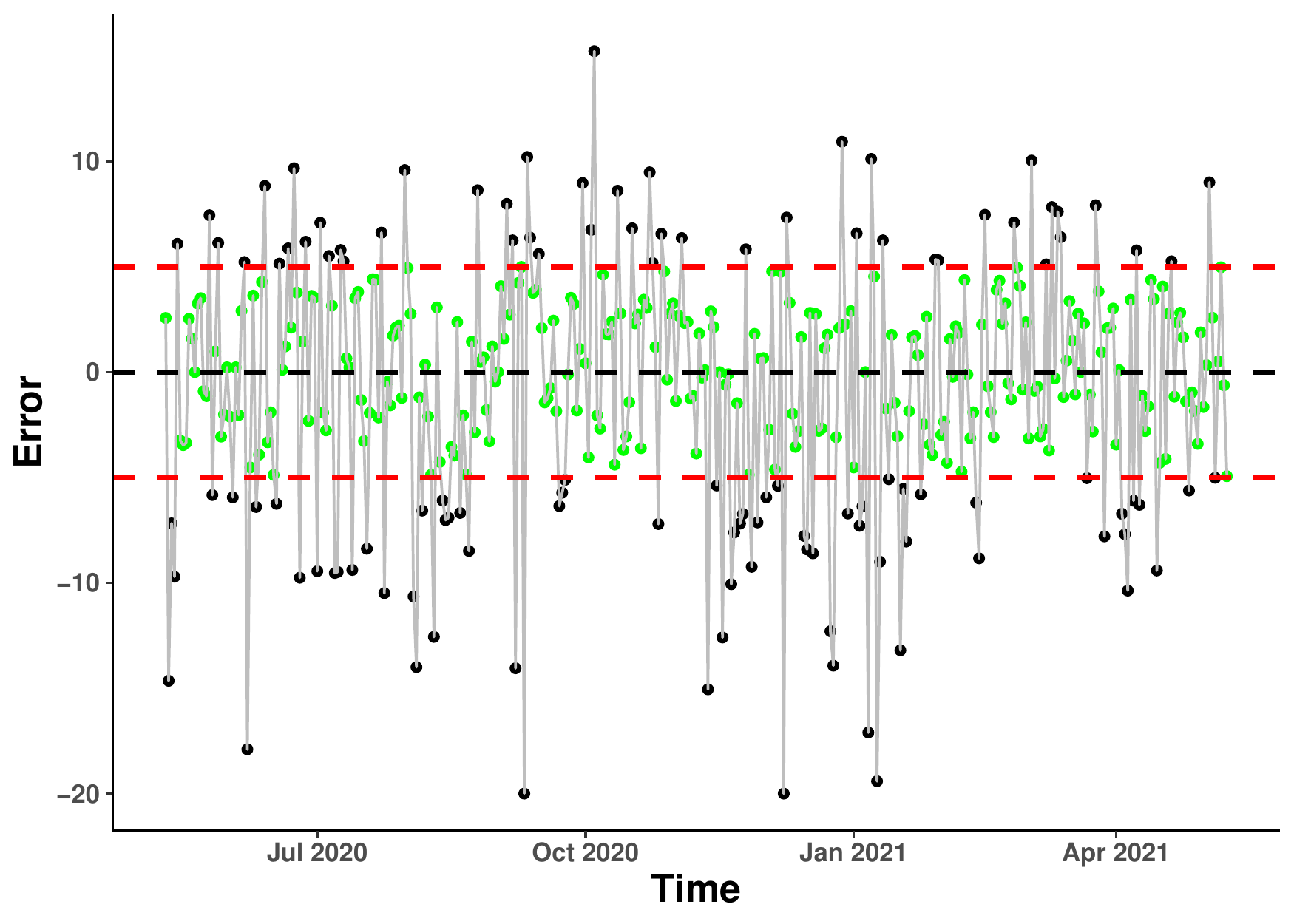}
	\caption{\label{fig:mse_1day_Metrop}}
	\end{subfigure}
	\caption{\emph{Metropolitan} model results: (a) Predictions one, two, five and seven days ahead across one year and relative true values; (b) Forecast errors $1$ day ahead, the predictions inside the two black dotted horizontal lines have absolute error below $5\%$.}
\end{figure}

\begin{figure}[tb]
	\centering
	\begin{subfigure}[b]{0.4\textwidth}
	\includegraphics[width=\textwidth]{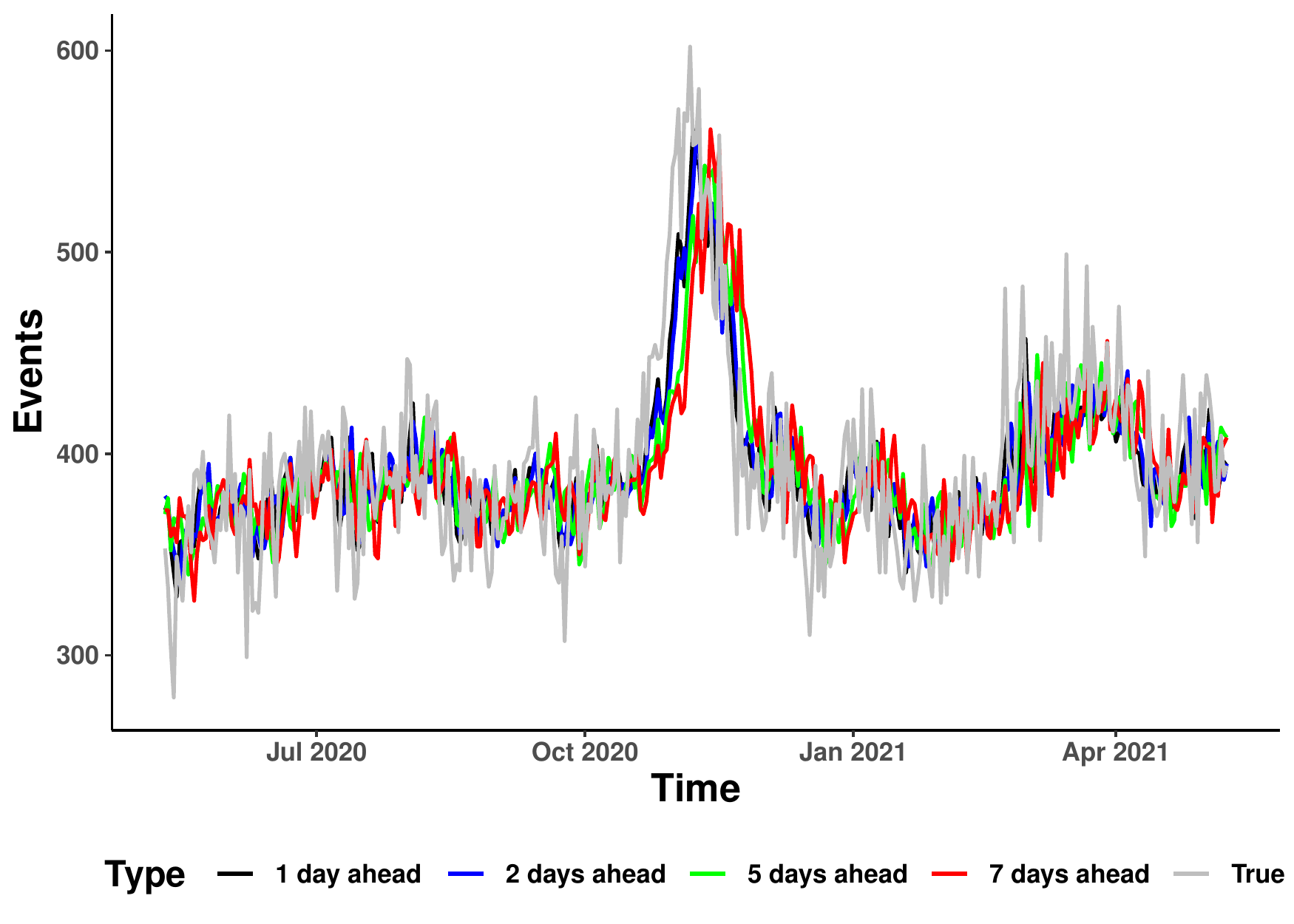}
	\caption{\label{fig:plot_totL}}
	\end{subfigure}
\begin{subfigure}[b]{0.45\textwidth}
	\centering
	\includegraphics[width=\textwidth]{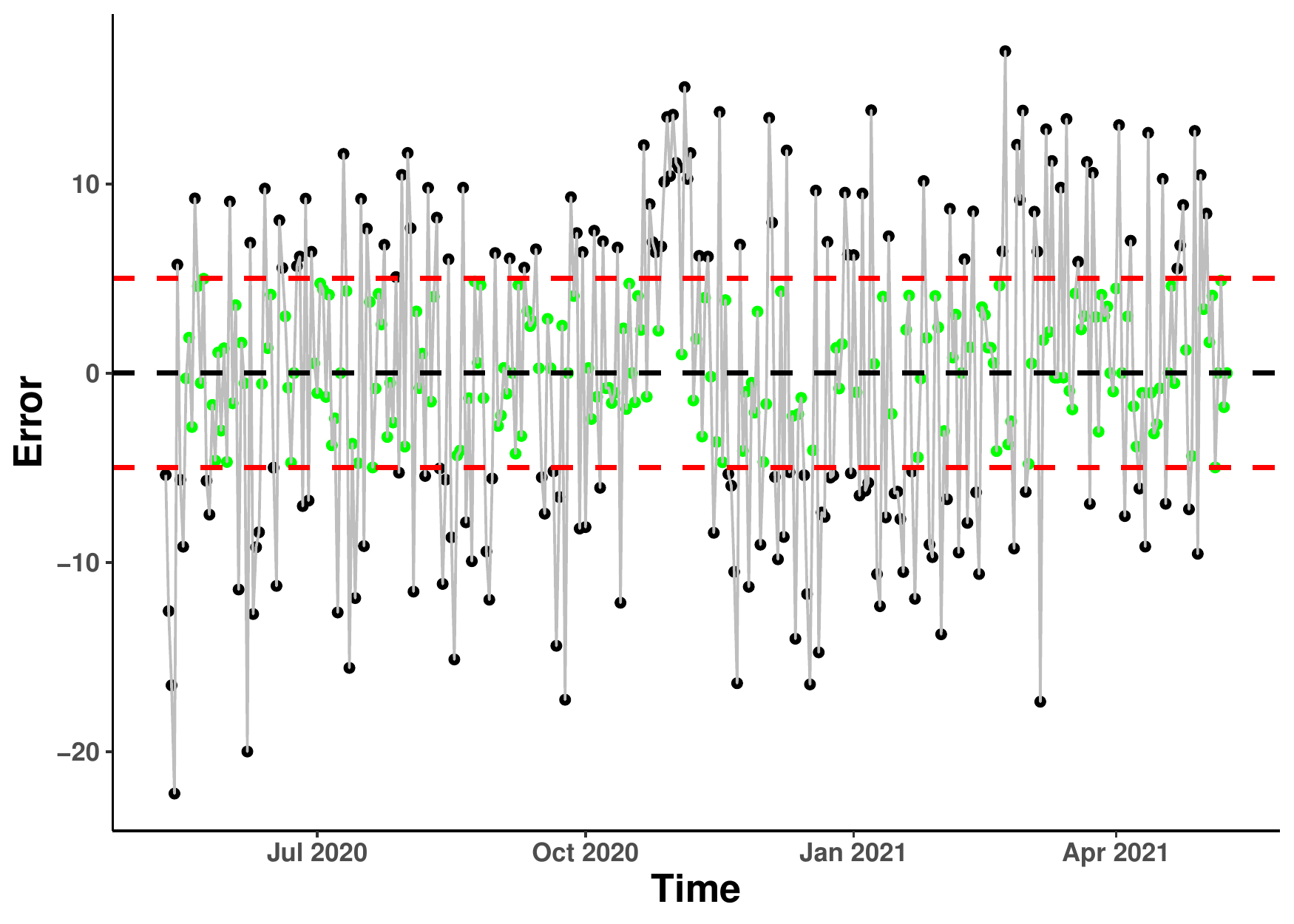}
	\caption{\label{fig:mse_1day_Laghi} }
	\end{subfigure}
		\caption{\emph{Lakes} model results: (a) Predictions one, two, five and seven days ahead across one year and relative true values; (b) Forecast errors $1$ day ahead, the predictions inside the two black dotted horizontal lines have absolute error below $5\%$.}
\end{figure}

\begin{figure}[tb]
	\centering
			\begin{subfigure}[b]{0.4\textwidth}
	\includegraphics[width=\textwidth]{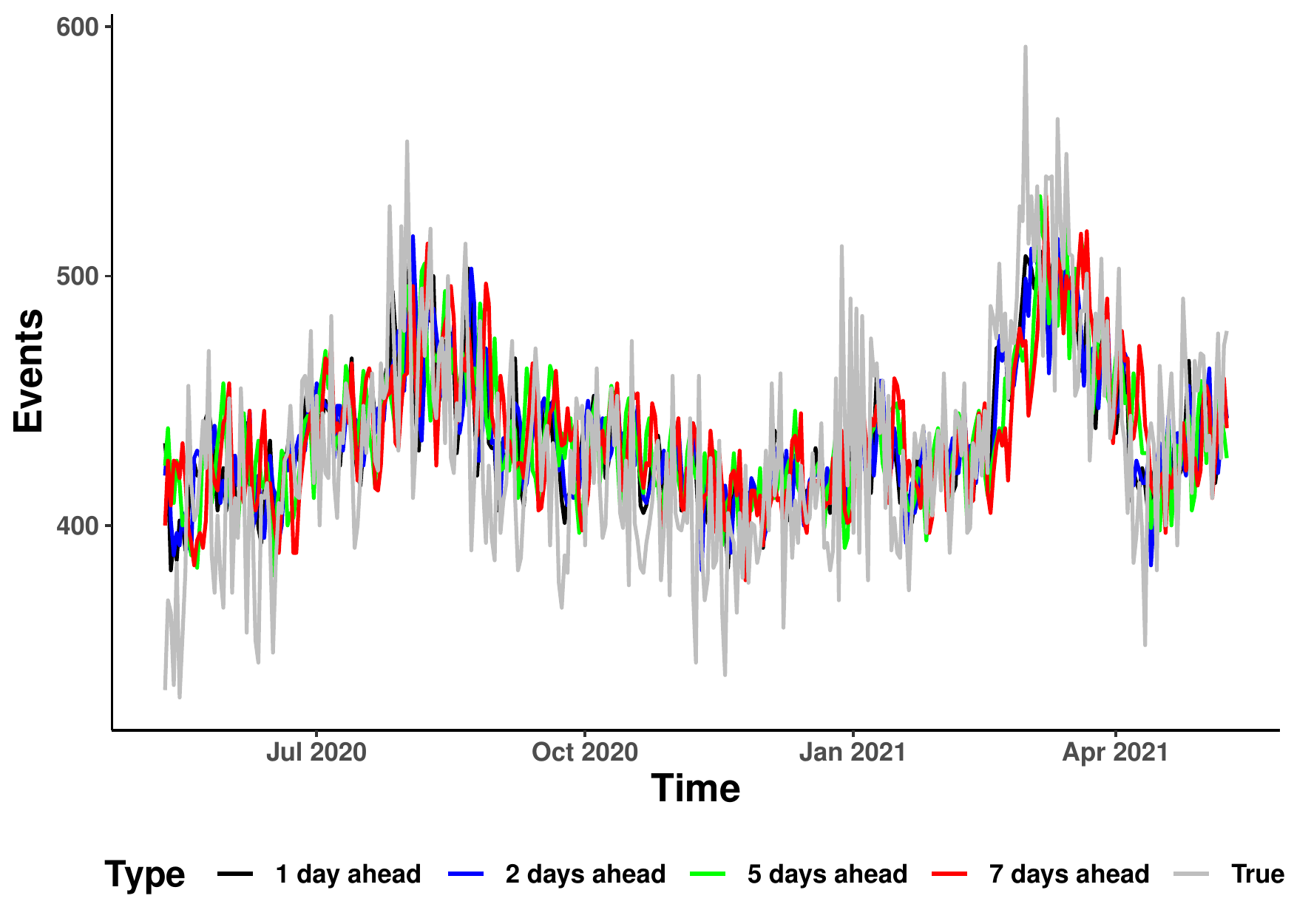}
	\caption{\label{fig:Predictions_Alpina} }
	\end{subfigure}
		\begin{subfigure}[b]{0.45\textwidth}
	\includegraphics[width=\textwidth]{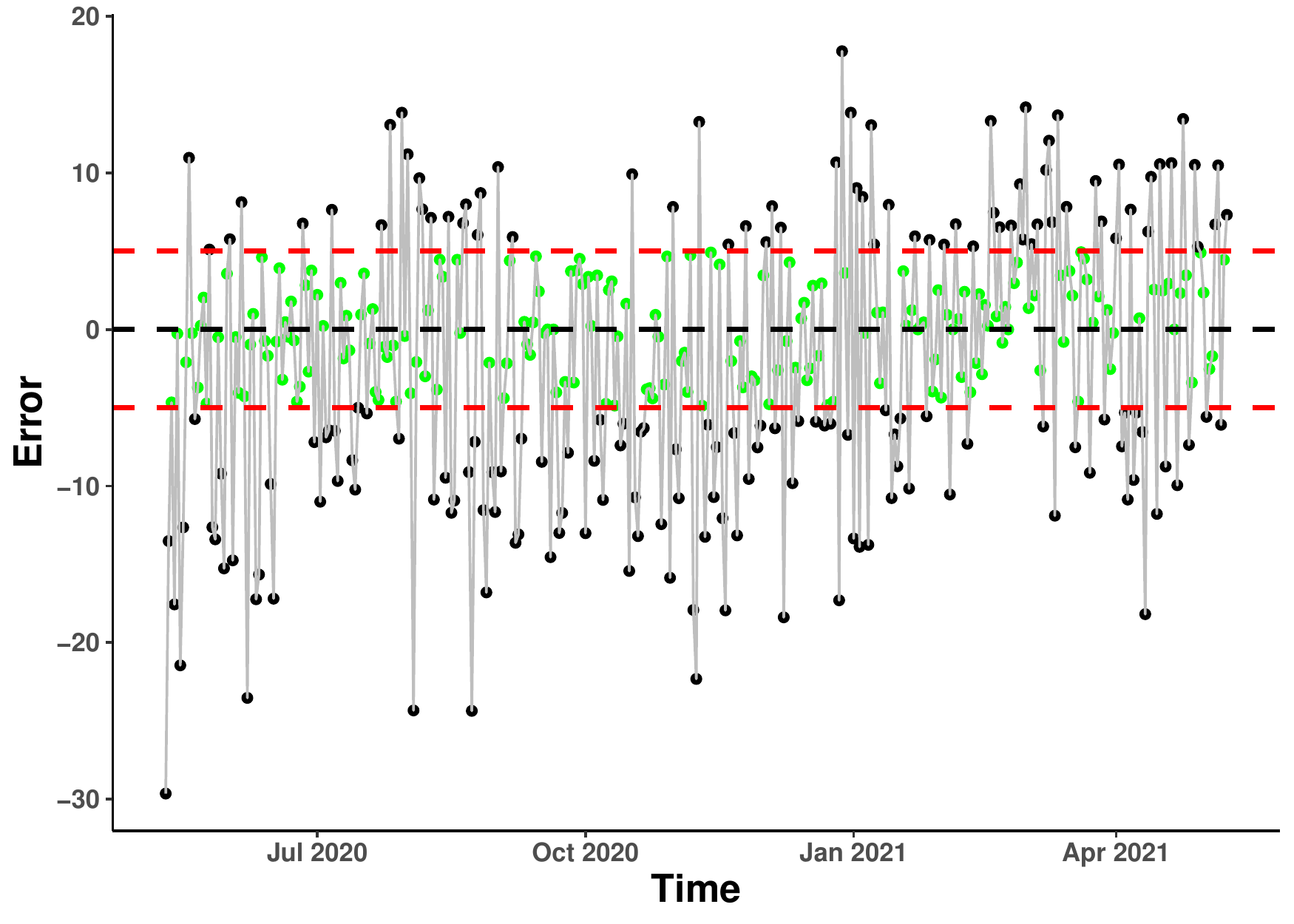}
	\caption{\label{fig:mse_1day_Alpina} }
	\end{subfigure}
		\caption{\emph{Alps} model results: (a) Predictions one, two, five and seven days ahead across one year and relative true values; (b) Forecast errors $1$ day ahead, the predictions inside the two black dotted horizontal lines have absolute error below $5\%$.}
\end{figure}

The model presented was selected considering the behavior of the \emph{Plain} SOREU. However, we also applied it to the other SOREUs. The one day ahead mean absolute error is still acceptable considering the \emph{Alps} ($6.241\%$) and \emph{Lakes} ($5.815\%$) SOREUs, while it equals $4.309\%$ if we forecast the \emph{Metropolitan} SOREU. This is probably due to the presence of strong irregular patterns in the SOREUs \emph{Alps}, and \emph{Lakes}. Looking at the \emph{Metropolitan} SOREU, Figure \ref{fig:plot_totM} shows the predictions one, two, five, and seven days ahead, while Figure \ref{fig:mse_1day_Metrop} describes the prediction errors if we forecast one day ahead. In the same way, Figures \ref{fig:plot_totL} and \ref{fig:mse_1day_Laghi} illustrate the results for the \emph{Lakes} SOREU, and Figures \ref{fig:Predictions_Alpina} and \ref{fig:mse_1day_Alpina} for the \emph{Alps} SOREU. The change of the effects fitting the model to pre and post-COVID data for the \emph{Metropolitan}, \emph{Lakes} and \emph{Alps} SOREUs are reported in the supplementary materials.

After selecting the GAM model, we do some benchmark analysis considering as benchmark the ARIMA model computed automatically by the \texttt{auto.arima} \texttt{R} function \citep{makridakis1997arma}, the deterministic model that considers the previous observation as prediction, and a generalised linear model (GLM) for time series of counts (tsglm) \citep{christou2014quasi}. Forecasting across one year (one day ahead), the ARIMA model returns a mean absolute error equal to $11.177\%$, the deterministic model yields a mean absolute error equal to $7.359\%$, while for the tsglm model the MAE equals $10.461\%$. We conclude that the proposed model outperforms these simple benchmark models.

\section{Discussion}\label{end}

This paper presents a valuable model to predict the number of events in the \emph{Plain} SOREU during the COVID-19 pandemic with a reasonable error compatible with the AREU request and useful for efficiently planning their emergency activities  from a socio-economical prospective. The model is able to capture the dramatic daily and seasonal variation emerged during the COVID-19 pandemic. Applying the model in the remaining SOREUs (i.e., \emph{Metropolitan}, \emph{Alps} and \emph{Lakes}) notable results are also obtained. The model proposed then appears to generalize to slightly different contexts. A future direction of research could be to apply this model using emergent data from other a future direction of research could then be to apply this model using emergent data from other regions of Italy or other countries.

Another further direction would be the application of the Generalized Additive Mixed Model (GAMM), i.e., an extension of GAM incorporating random effects. GAMM better deals with the data autocorrelation structure, however, at the price of a high computational cost \citep{lin1999inference}. In addition, some analysis regarding interaction terms might be developed, taking care of possible over-fitting. Finally, a Bayesian extension might also be helpful by assigning to covariates appropriate Markov random field priors with different forms and degrees of smoothness to deal with the trend and seasonal components \citep{fahrmeir2001bayesian}.

\section*{Acknowledgment}
This research was supported by the Lombardy region in Italy (DGR N. XI/3017/2020, ``TSUNAMI: Prediction of the Impact of the COVID-19 Wave on the Lombardy Emergency and Urgency System''). This funded research projects in the health sector related to the COVID-19 emergency.

\section*{Author contributions}

\textbf{Angela Andreella}: conceptualization, software, data curation, formal analysis, investigation, and writing - original draft. \textbf{Antonietta Mira}: conceptualization, funding acquisition, project administration, writing - review \& editing, and supervision. \textbf{Spyros Balafas}: conceptualization, data curation, software, investigation, writing - review \& editing. \textbf{Ernst C. Wit}: conceptualization, writing - review \& editing, and supervision. \textbf{Fabrizio Ruggeri}: conceptualization, writing - review \& editing, and supervision. \textbf{Giovanni Nattino}: conceptualization, writing - review \& editing, and supervision. \textbf{Giulia Ghilardi}: supervision. \textbf{Guido Bertolini}: conceptualization, funding acquisition, writing - review \& editing, and supervision.

\section*{Declaration of Competing Interest}
The authors declare no competing interests.

\clearpage

\bibliographystyle{apalike}
\bibliography{bibliography}
\clearpage

\title{SUPPLEMENTARY MATERIALS for ``A predictive model for planning emergency events rescue during COVID-19 in Lombardy, Italy''}

\date{}

\maketitle

We show here the plots of the component smooth functions for \emph{Metropolitan}, \emph{Lakes} and \emph{Alps} SOREUs using the full dataset and data pre-COVID pandemic era.

\section{Metropolitan}

\begin{figure}[H]
	\centering
	\begin{subfigure}[b]{0.4\textwidth}
		\includegraphics[width=\textwidth]{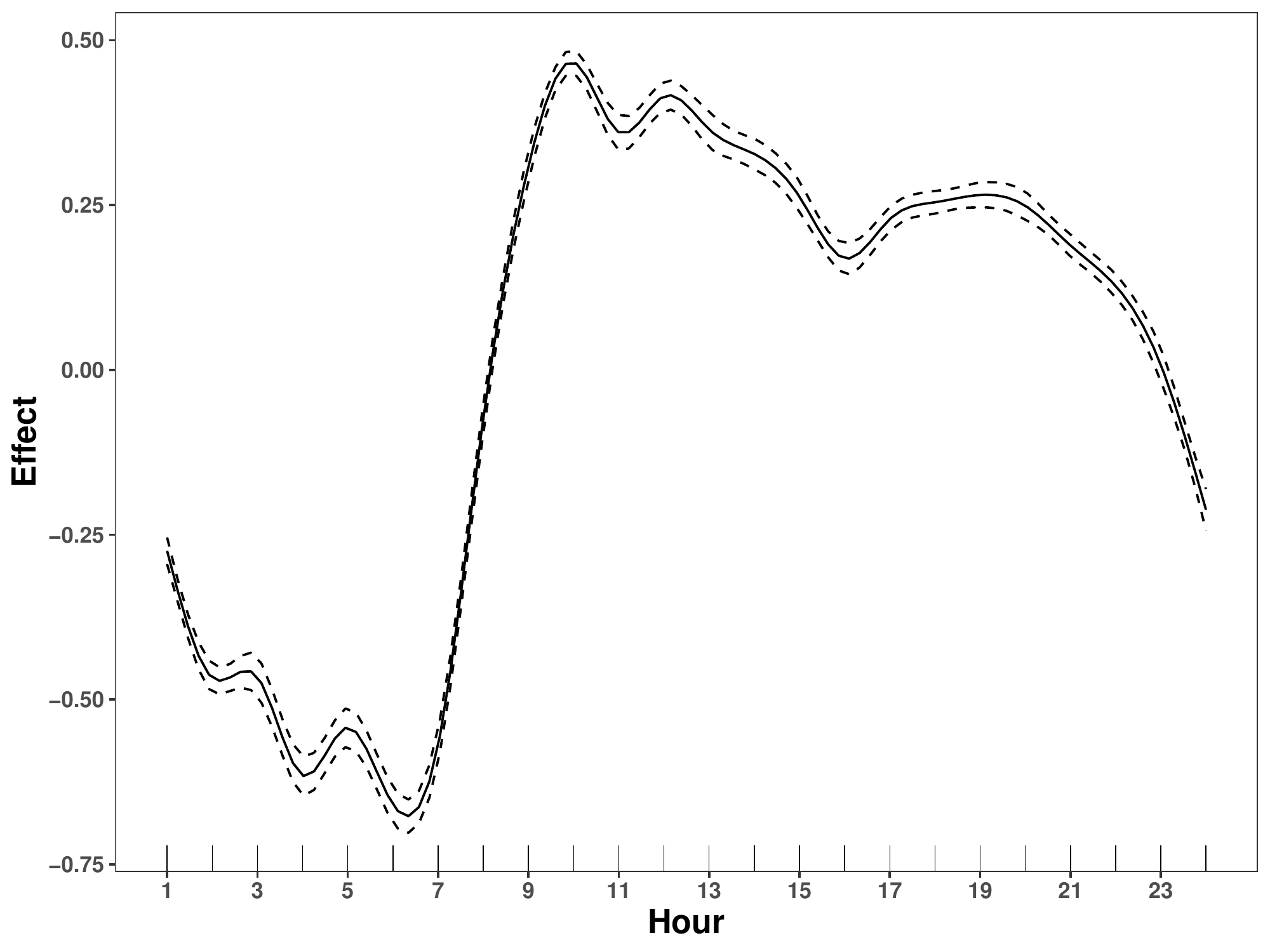}
		\caption{\label{fig1}}
	\end{subfigure}
	\begin{subfigure}[b]{0.4\textwidth}
		\includegraphics[width=\textwidth]{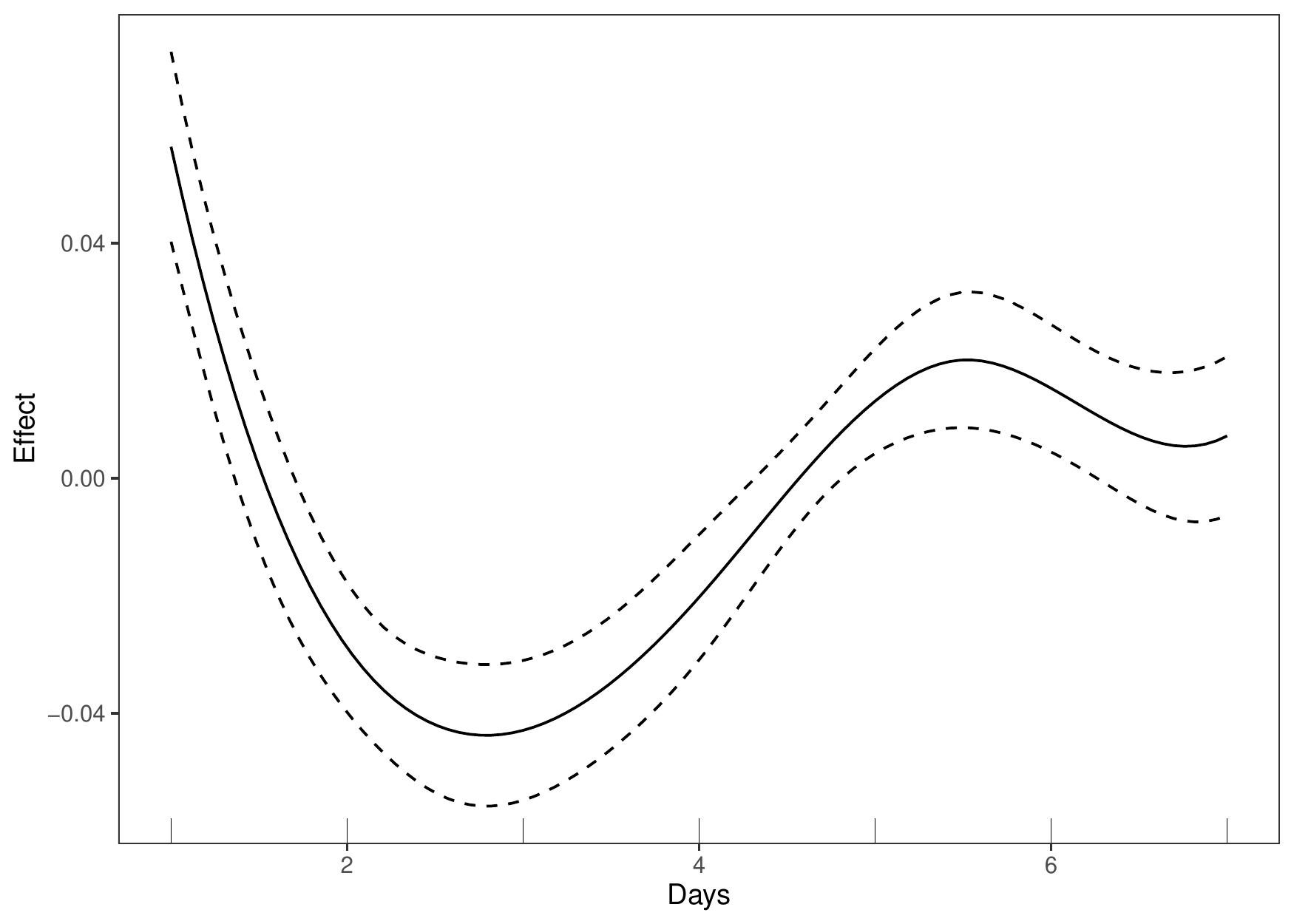}
		\caption{\label{fig11}}
	\end{subfigure}
	\begin{subfigure}[b]{0.4\textwidth}
		\includegraphics[width=\textwidth]{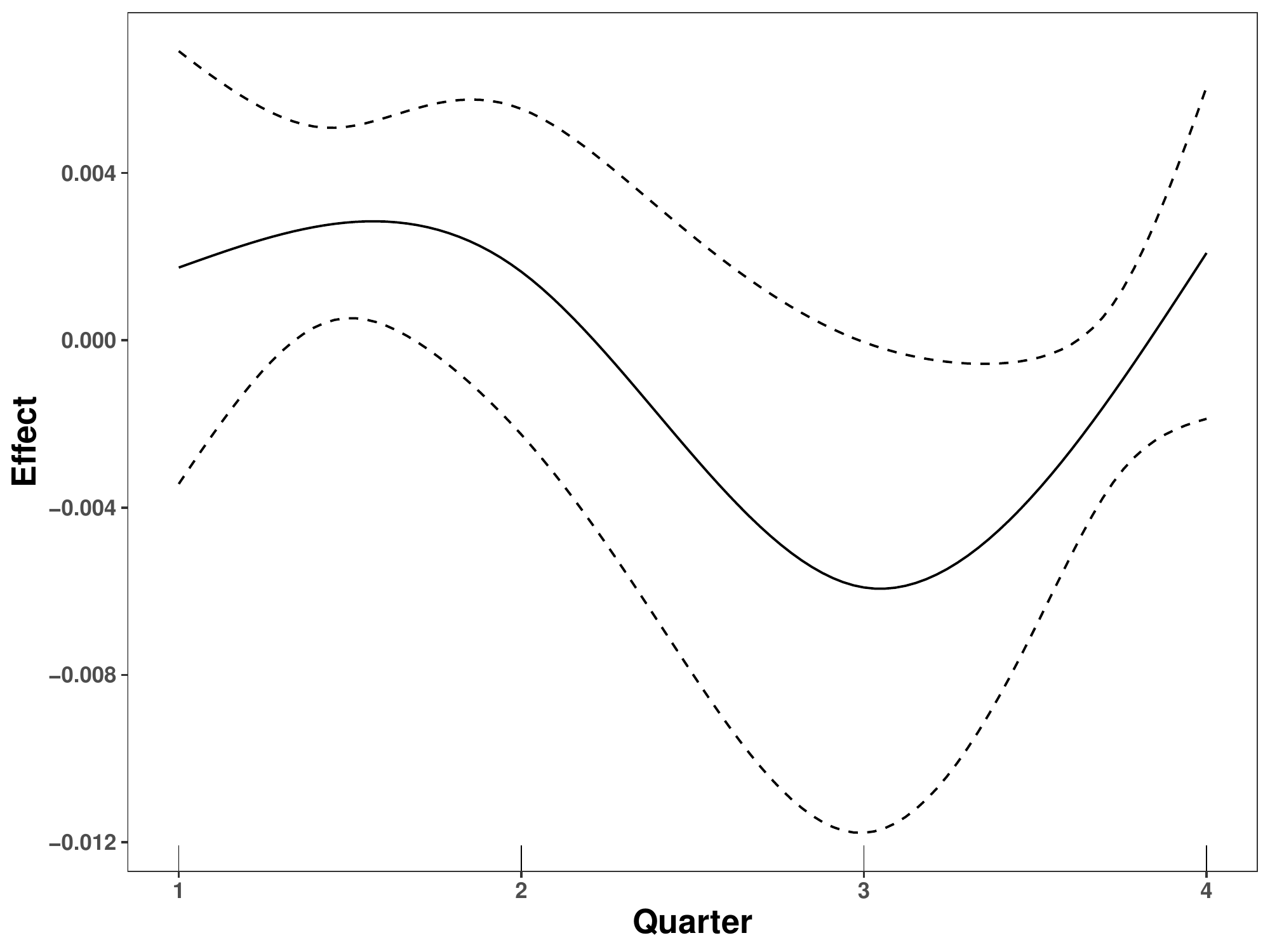}
		\caption{\label{fig2}}
	\end{subfigure}
	\begin{subfigure}[b]{0.4\textwidth}
		\includegraphics[width=\textwidth]{Figures/te_Pianura.pdf}
		\caption{\label{fig21}}
	\end{subfigure}
\caption{\emph{Metropolitan} model coefficients plots during the COVID-19 pandemic: (a) Effect of the hour of the day; (b) Effect of the day of the week, where $1$ stands for Monday; (c) Effect of the quarters of the year, where $1$ stands for the first quarter of the year; (d) Effect of the interaction between days and hours, where $1$ stands for Monday.}
\label{fig:coef}

\end{figure}

\begin{figure}[H]
	\centering
	\begin{subfigure}[b]{0.4\textwidth}
		\includegraphics[width=\textwidth]{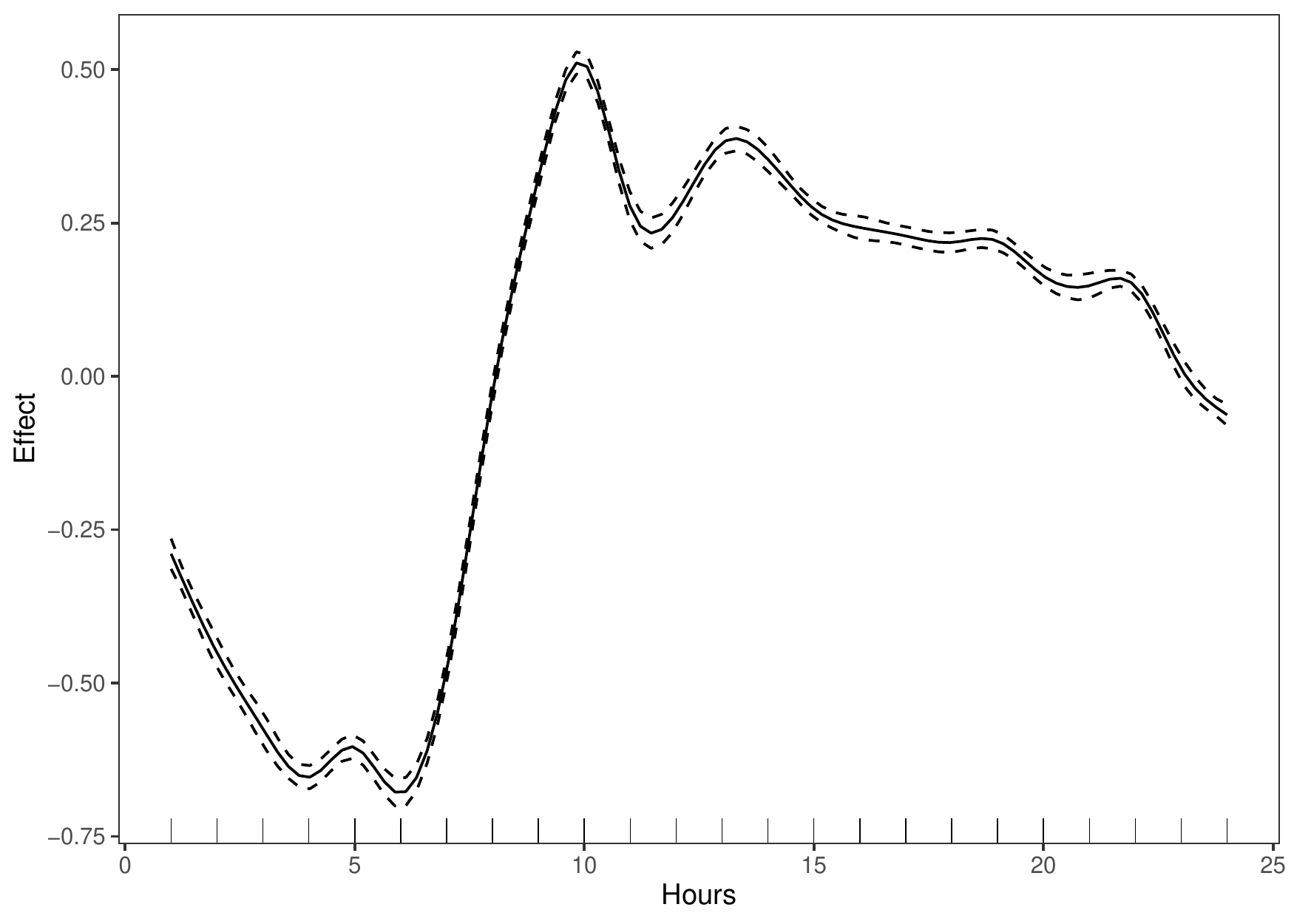}
		\caption{\label{fig1pre}}
		
	\end{subfigure}
	\begin{subfigure}[b]{0.4\textwidth}
		\includegraphics[width=\textwidth]{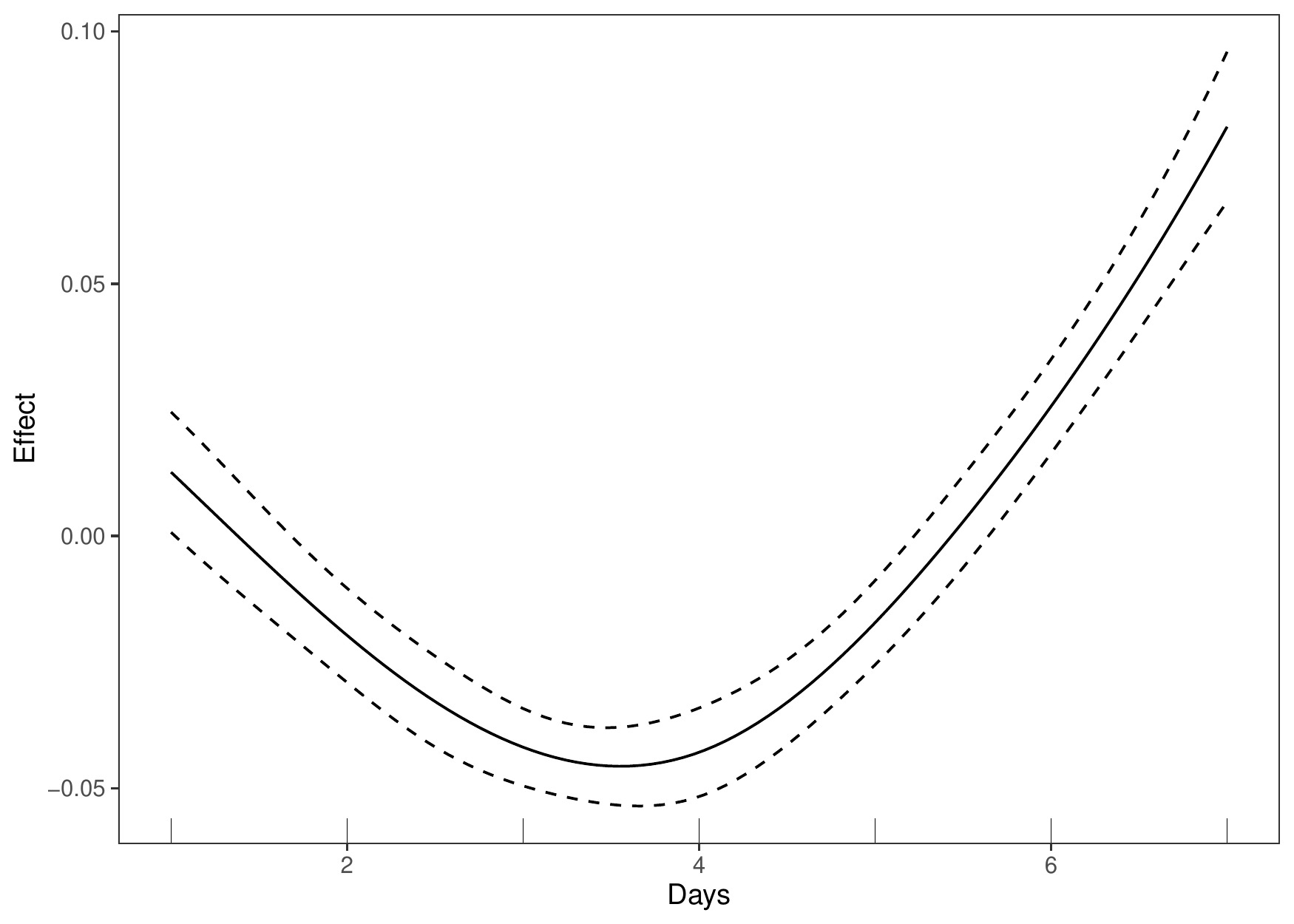}
		\caption{
		\label{fig11pre}}
	\end{subfigure}
	\begin{subfigure}[b]{0.4\textwidth}
		\includegraphics[width=\textwidth]{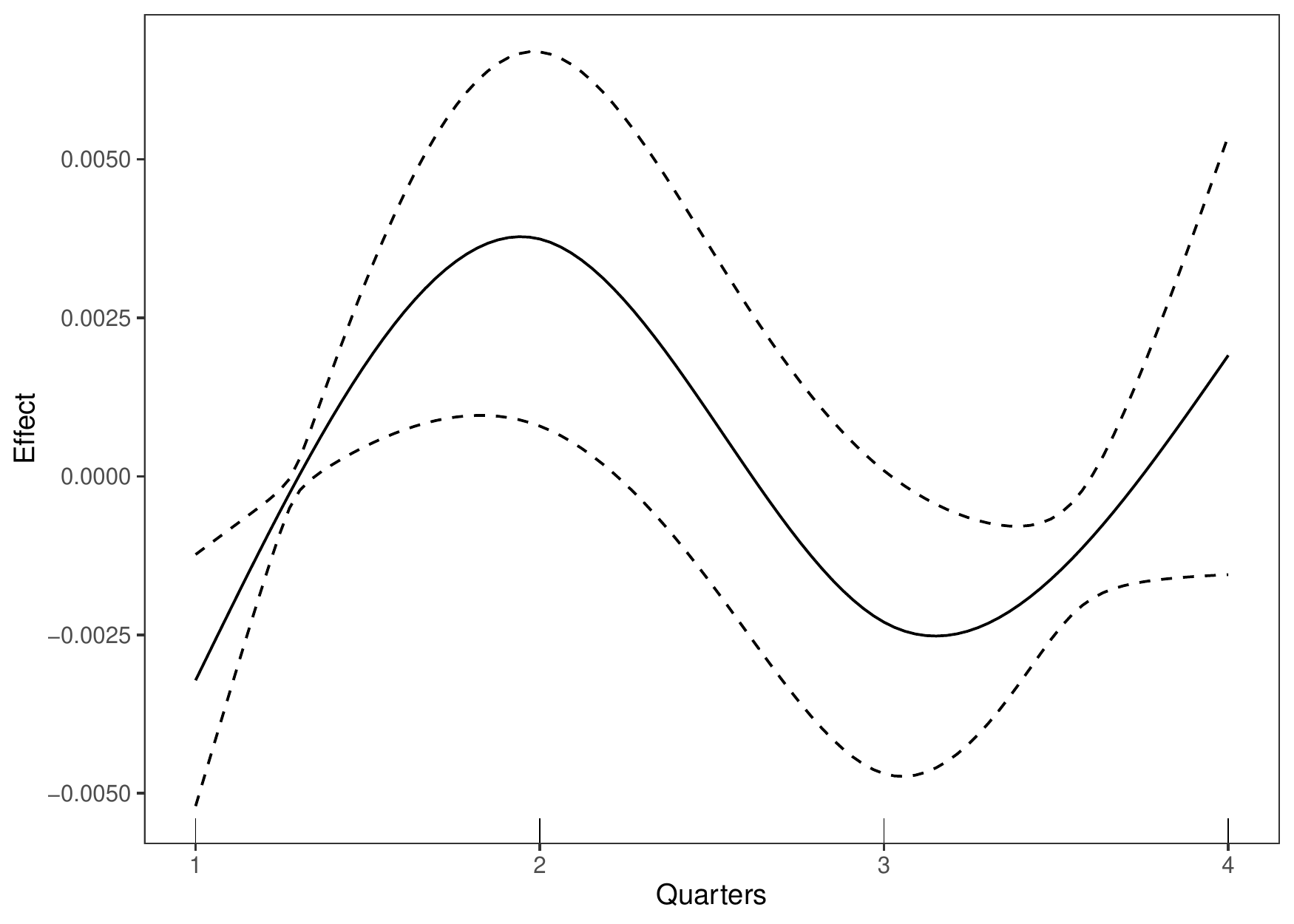}
		\caption{\label{fig2pre}}
	\end{subfigure}
	\begin{subfigure}[b]{0.4\textwidth}
		\includegraphics[width=\textwidth]{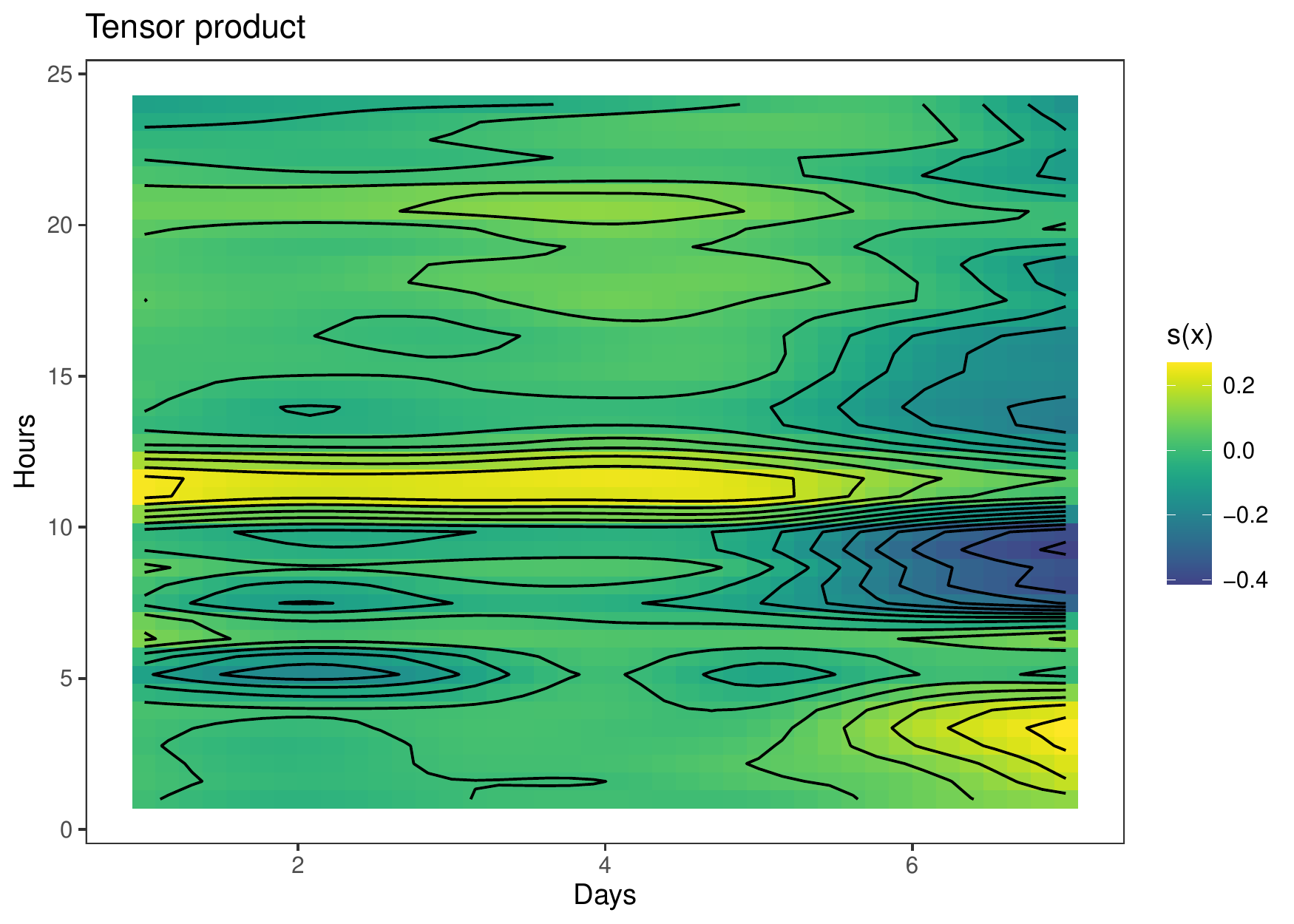}
		\caption{\label{fig21pre}}
	\end{subfigure}
\caption{\emph{Metropolitan} model coefficients plots before the COVID-19 pandemic: (a) Effect of the hour of the day; (b) Effect of the day of the week, where $1$ stands for Monday; (c) Effect of the quarters of the year, where $1$ stands for the first quarter of the year; (d) Effect of the interaction between days and hours, where $1$ stands for Monday.}
\label{coef_preCovid}

\end{figure}

\section{Lakes}

\begin{figure}[H]
	\centering
	\begin{subfigure}[b]{0.4\textwidth}
		\includegraphics[width=\textwidth]{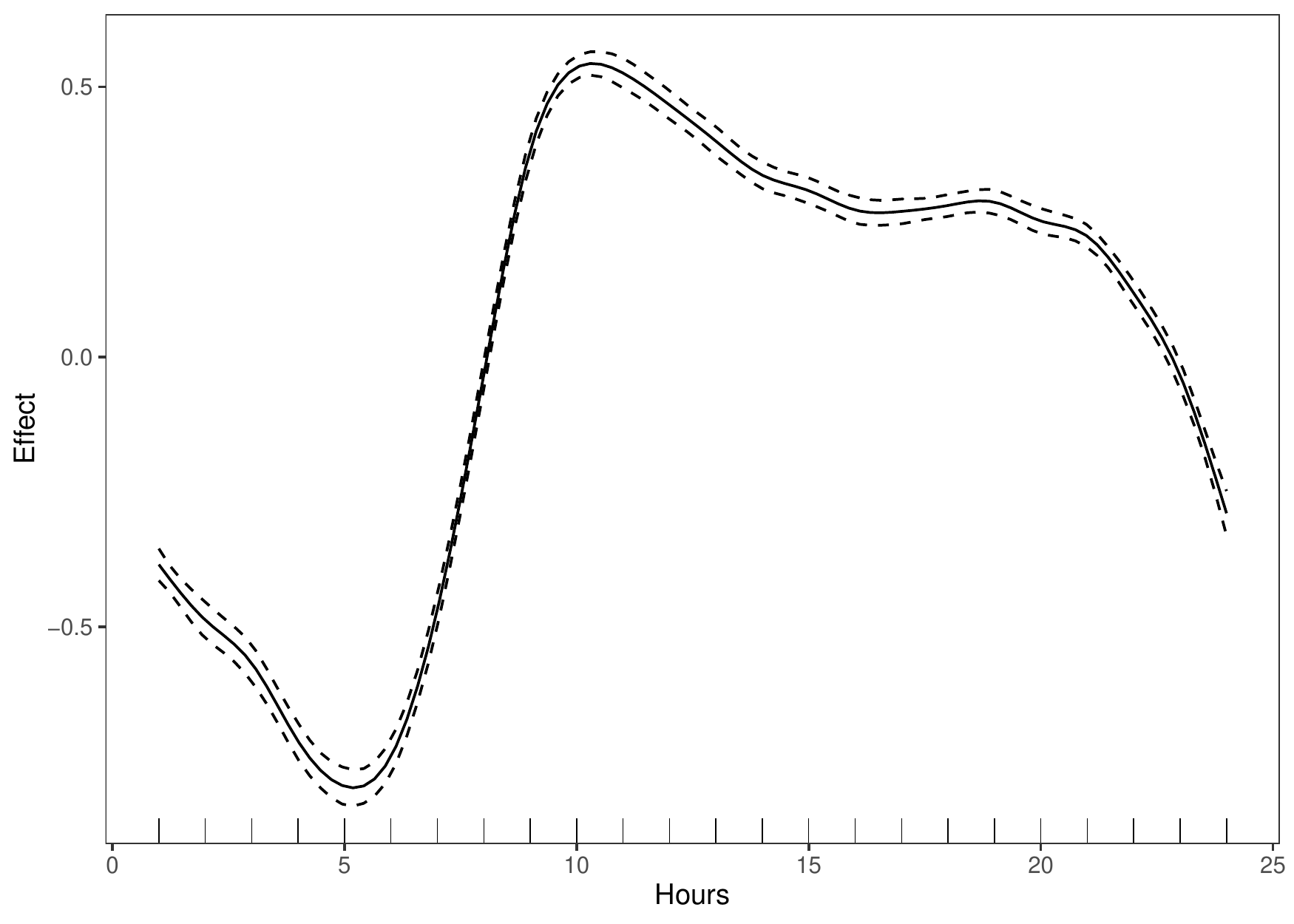}
		\caption{\label{fig1}}
		
	\end{subfigure}
	\begin{subfigure}[b]{0.4\textwidth}
		\includegraphics[width=\textwidth]{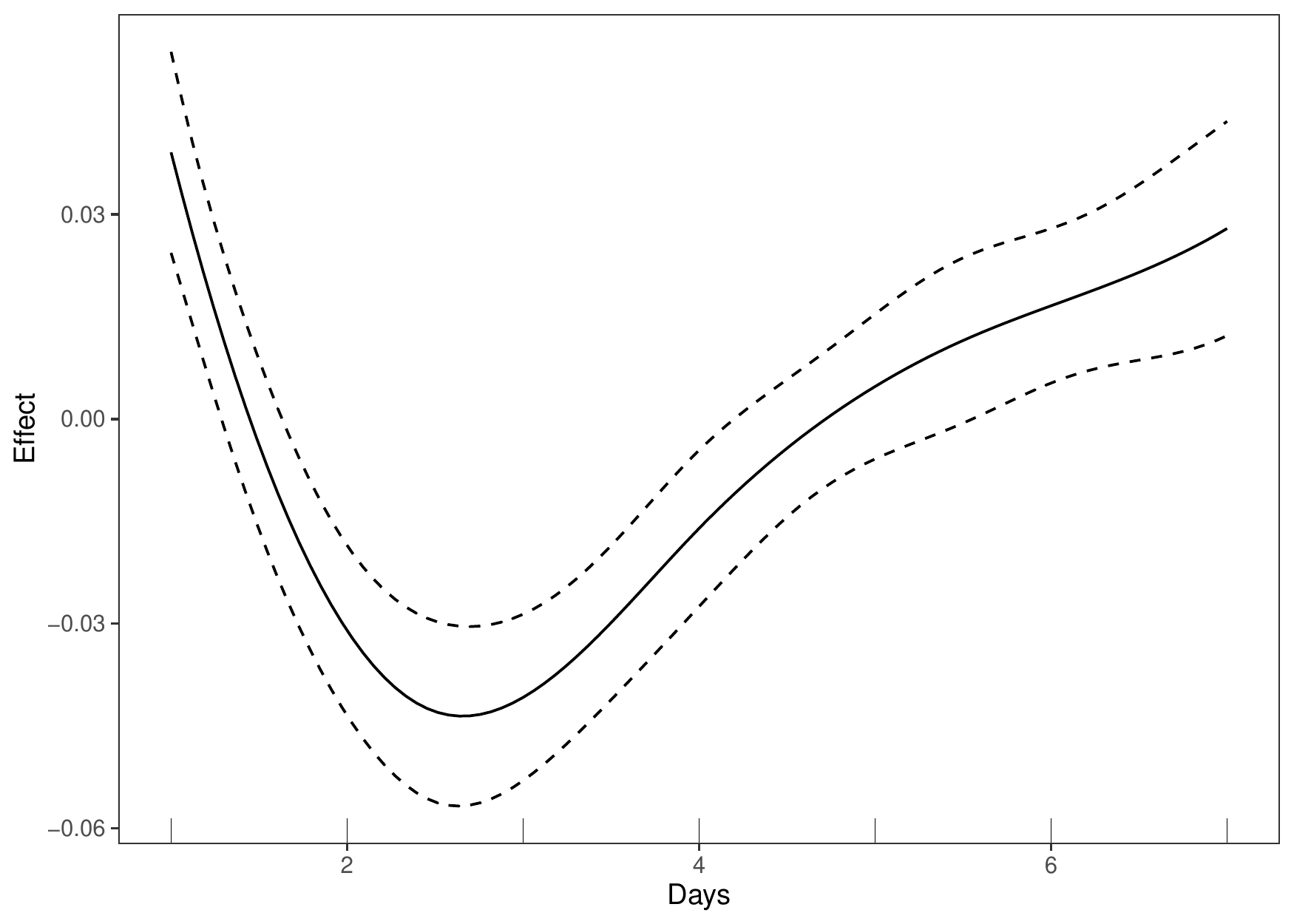}
		\caption{\label{fig11}}
		
	\end{subfigure}
	\begin{subfigure}[b]{0.4\textwidth}
		\includegraphics[width=\textwidth]{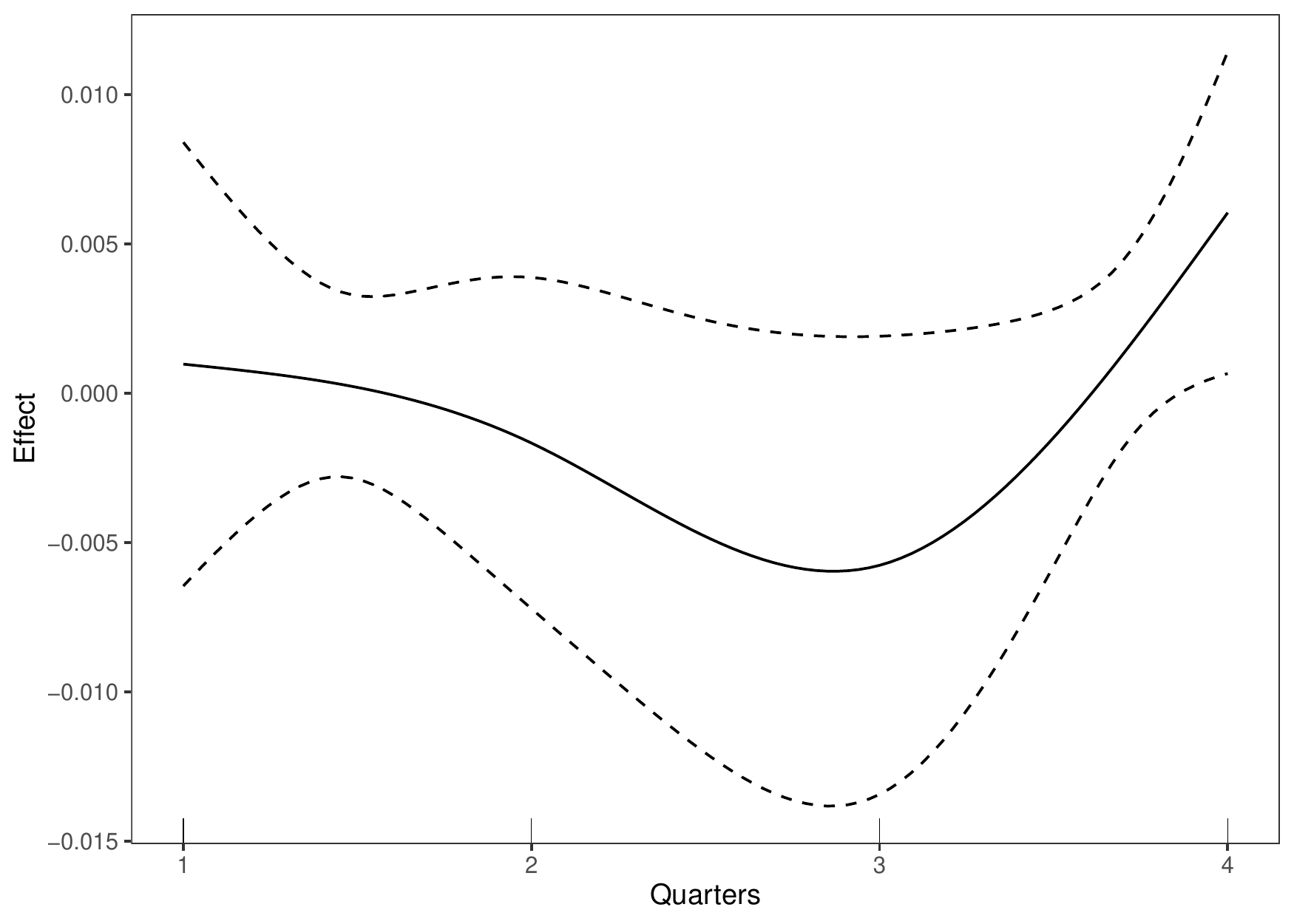}
		\caption{\label{fig2}}
		
	\end{subfigure}
	\begin{subfigure}[b]{0.4\textwidth}
		\includegraphics[width=\textwidth]{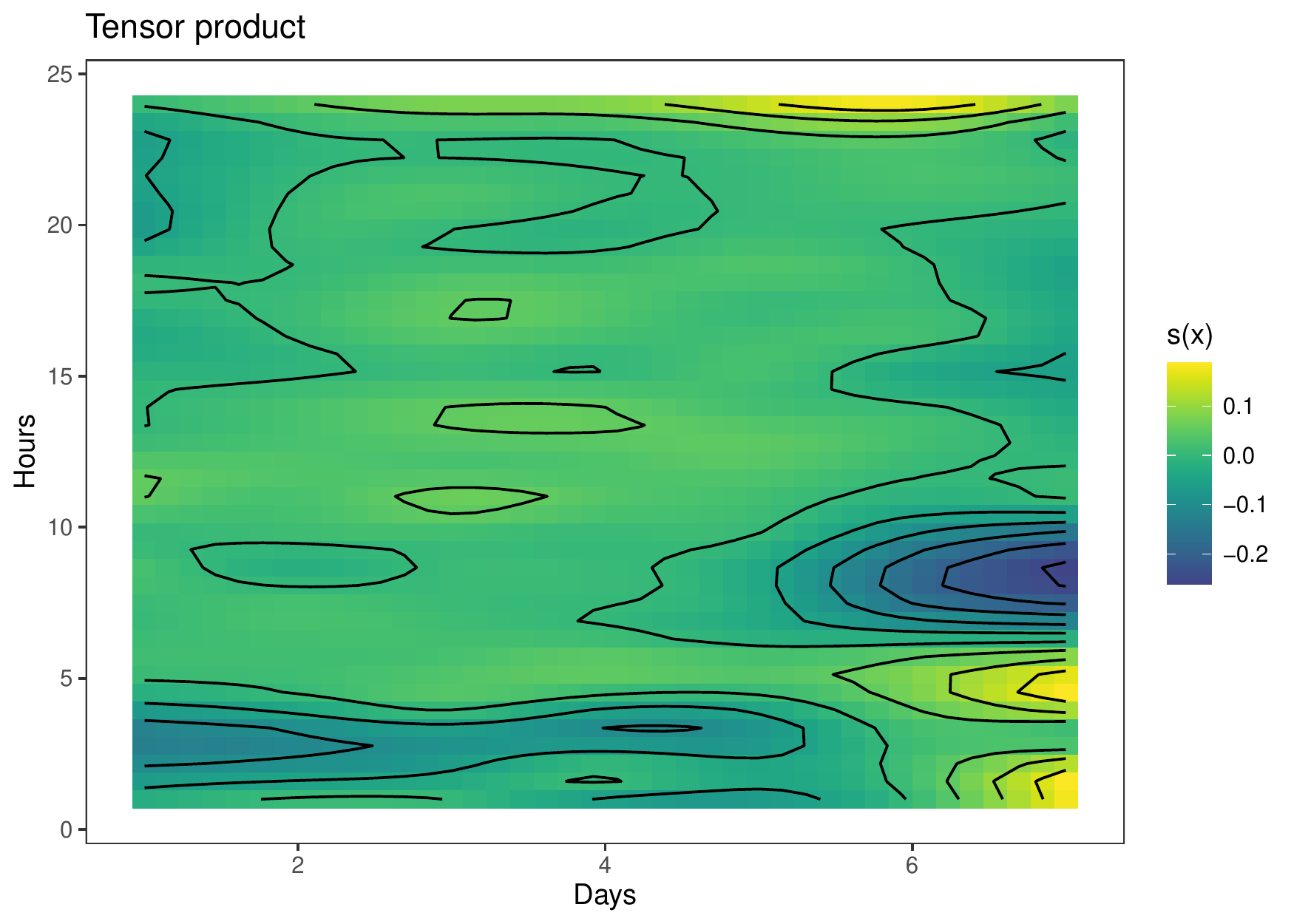}
		\caption{\label{fig21}}
		
	\end{subfigure}
\caption{\emph{Lakes} model coefficients plots during the COVID-19 pandemic: (a) Effect of the hour of the day; (b) Effect of the day of the week, where $1$ stands for Monday; (c) Effect of the quarters of the year, where $1$ stands for the first quarter of the year; (d) Effect of the interaction between days and hours, where $1$ stands for Monday.}
\label{fig:coef}

\end{figure}

\begin{figure}[H]
	\centering
	\begin{subfigure}[b]{0.4\textwidth}
		\includegraphics[width=\textwidth]{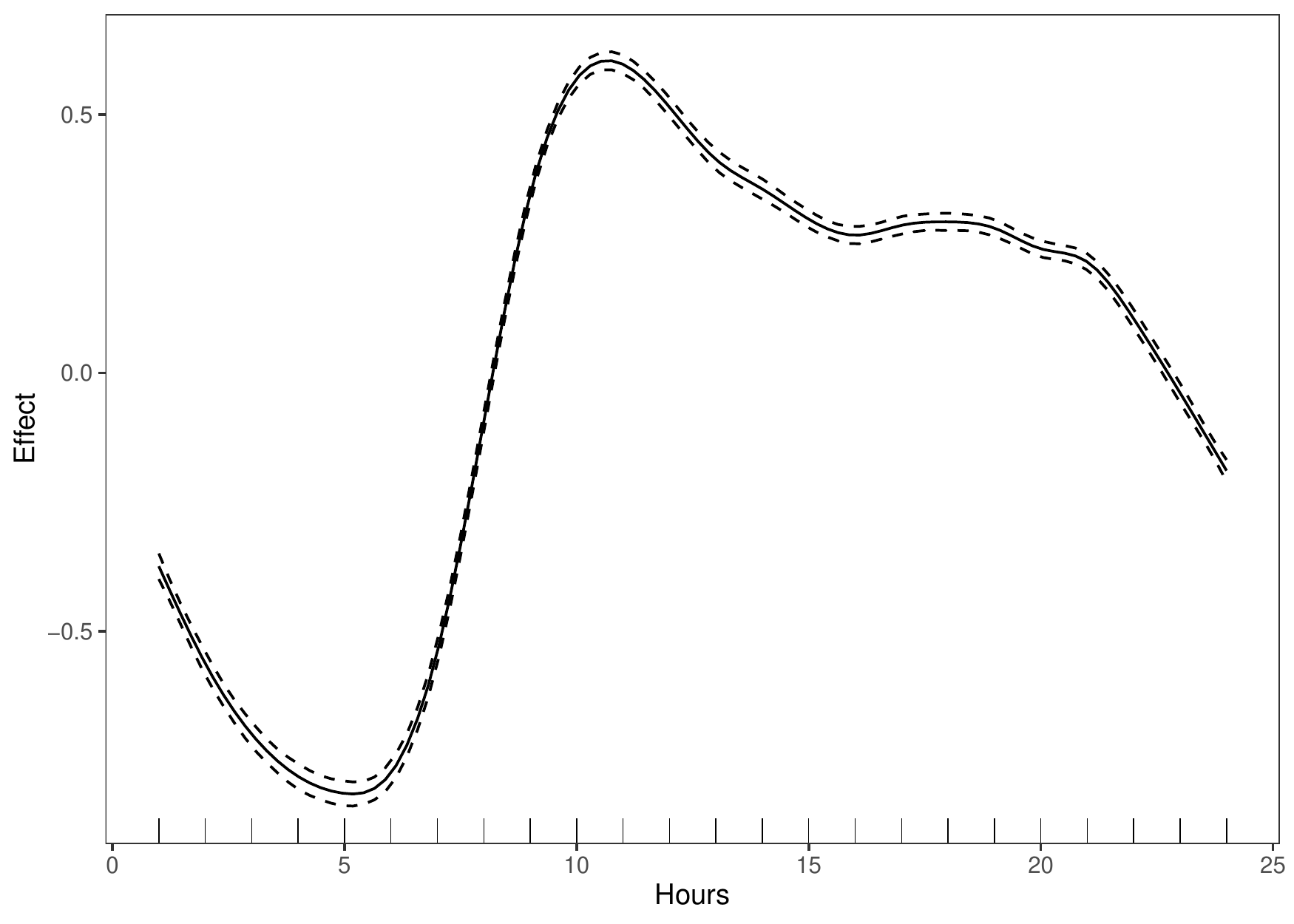}
		\caption{\label{fig1pre}}
		
	\end{subfigure}
	\begin{subfigure}[b]{0.4\textwidth}
		\includegraphics[width=\textwidth]{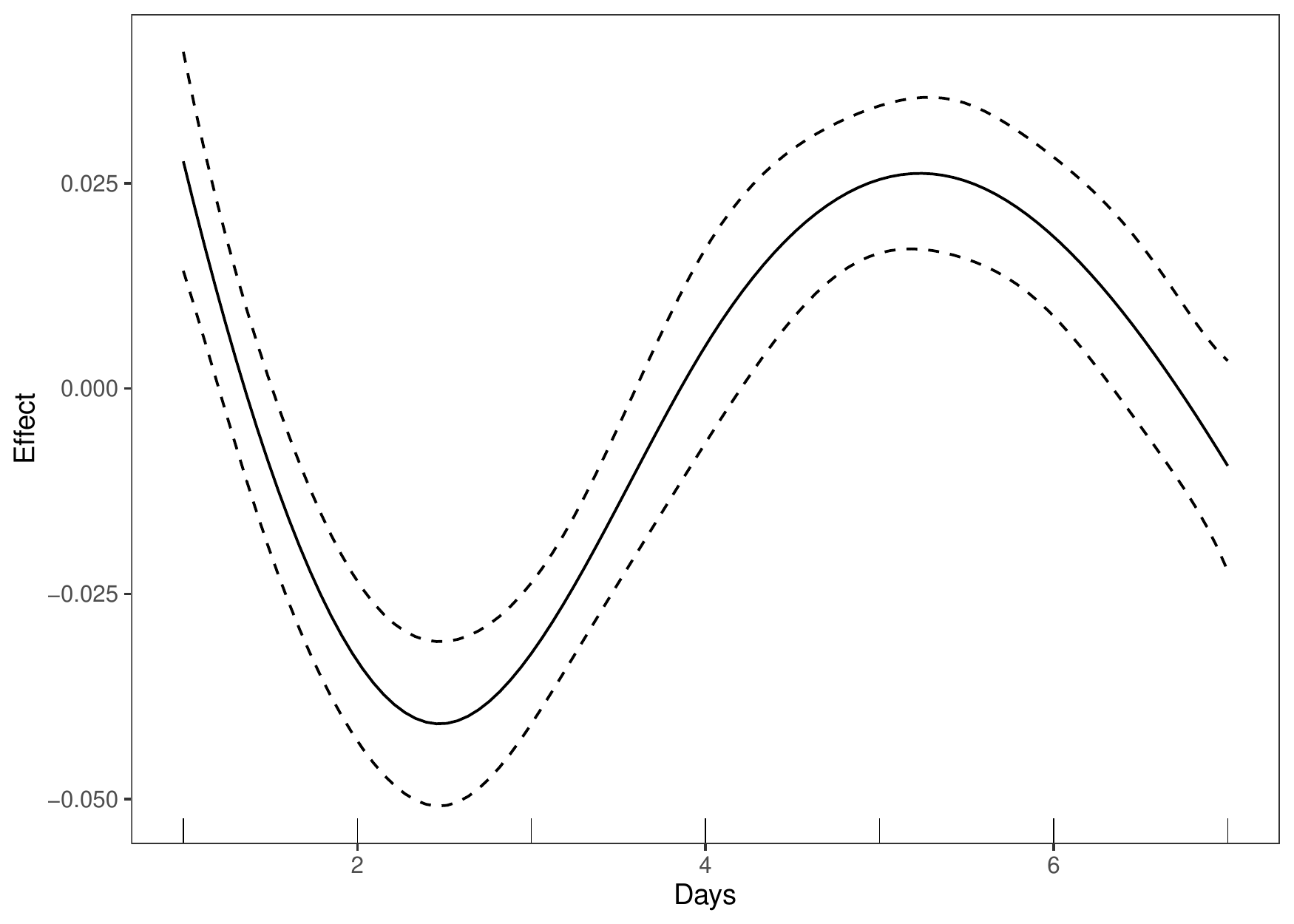}
		\caption{\label{fig11pre}}
		
	\end{subfigure}
	\begin{subfigure}[b]{0.4\textwidth}
		\includegraphics[width=\textwidth]{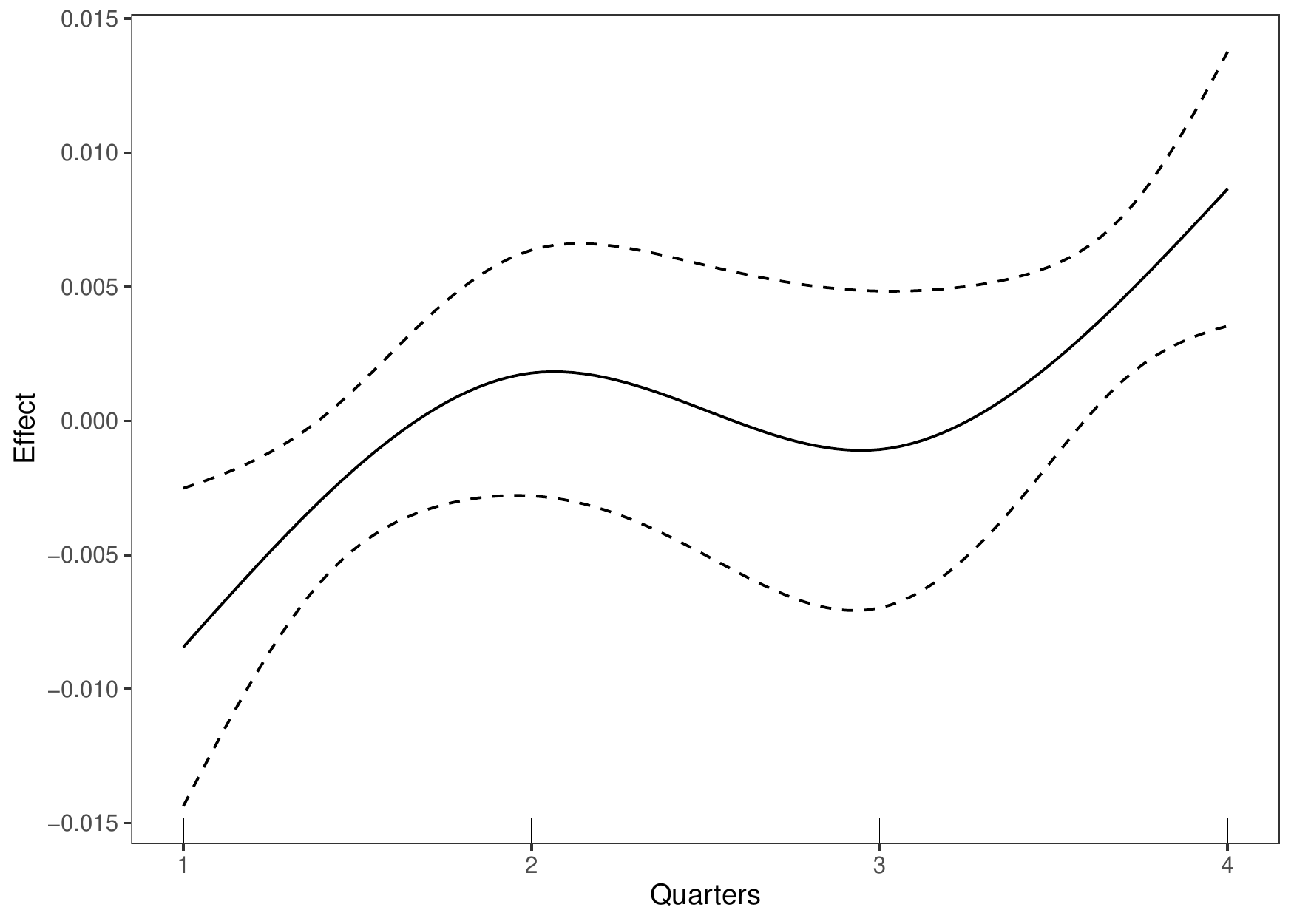}
		\caption{	\label{fig2pre}}
	
	\end{subfigure}
	\begin{subfigure}[b]{0.4\textwidth}
		\includegraphics[width=\textwidth]{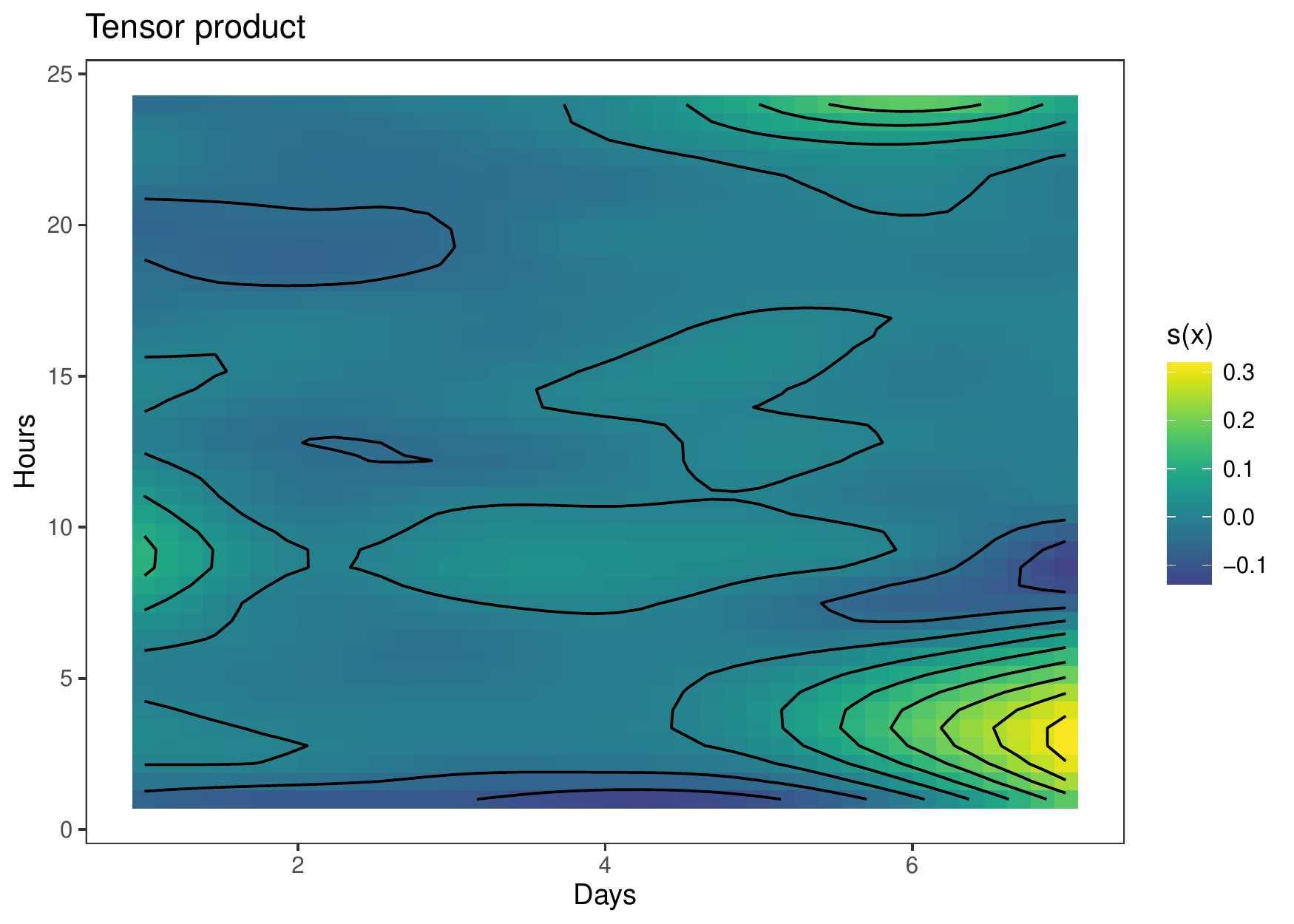}
		\caption{\label{fig21pre}}
		
	\end{subfigure}
\caption{\emph{Lakes} model coefficients plots before the COVID-19 pandemic: (a) Effect of the hour of the day; (b) Effect of the day of the week, where $1$ stands for Monday; (c) Effect of the quarters of the year, where $1$ stands for the first quarter of the year; (d) Effect of the interaction between days and hours, where $1$ stands for Monday.}
\label{coef_preCovid}

\end{figure}

\section{Alps}

\begin{figure}[H]
	\centering
	\begin{subfigure}[b]{0.4\textwidth}
		\includegraphics[width=\textwidth]{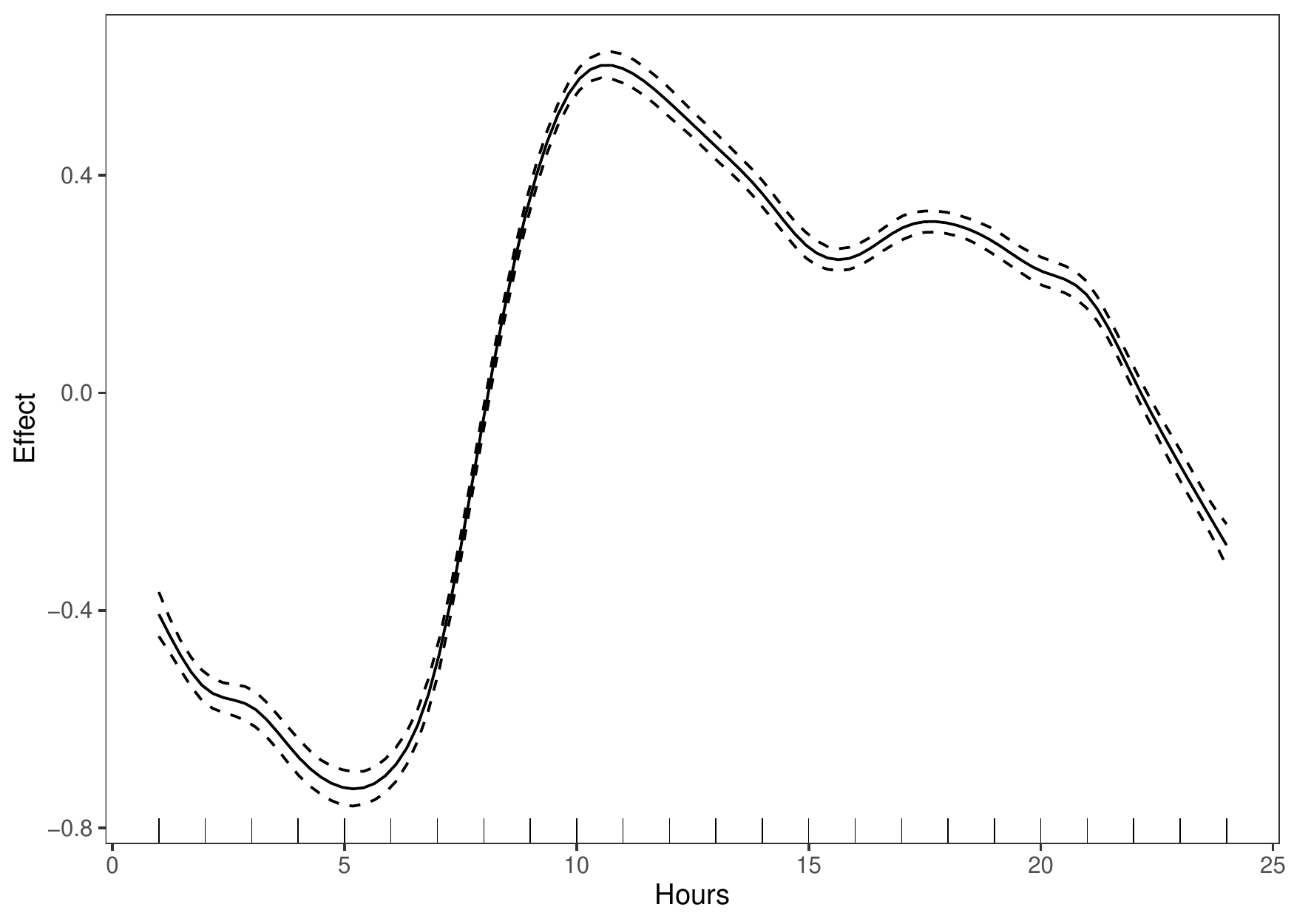}
		\caption{\label{fig1}}
		
	\end{subfigure}
	\begin{subfigure}[b]{0.4\textwidth}
		\includegraphics[width=\textwidth]{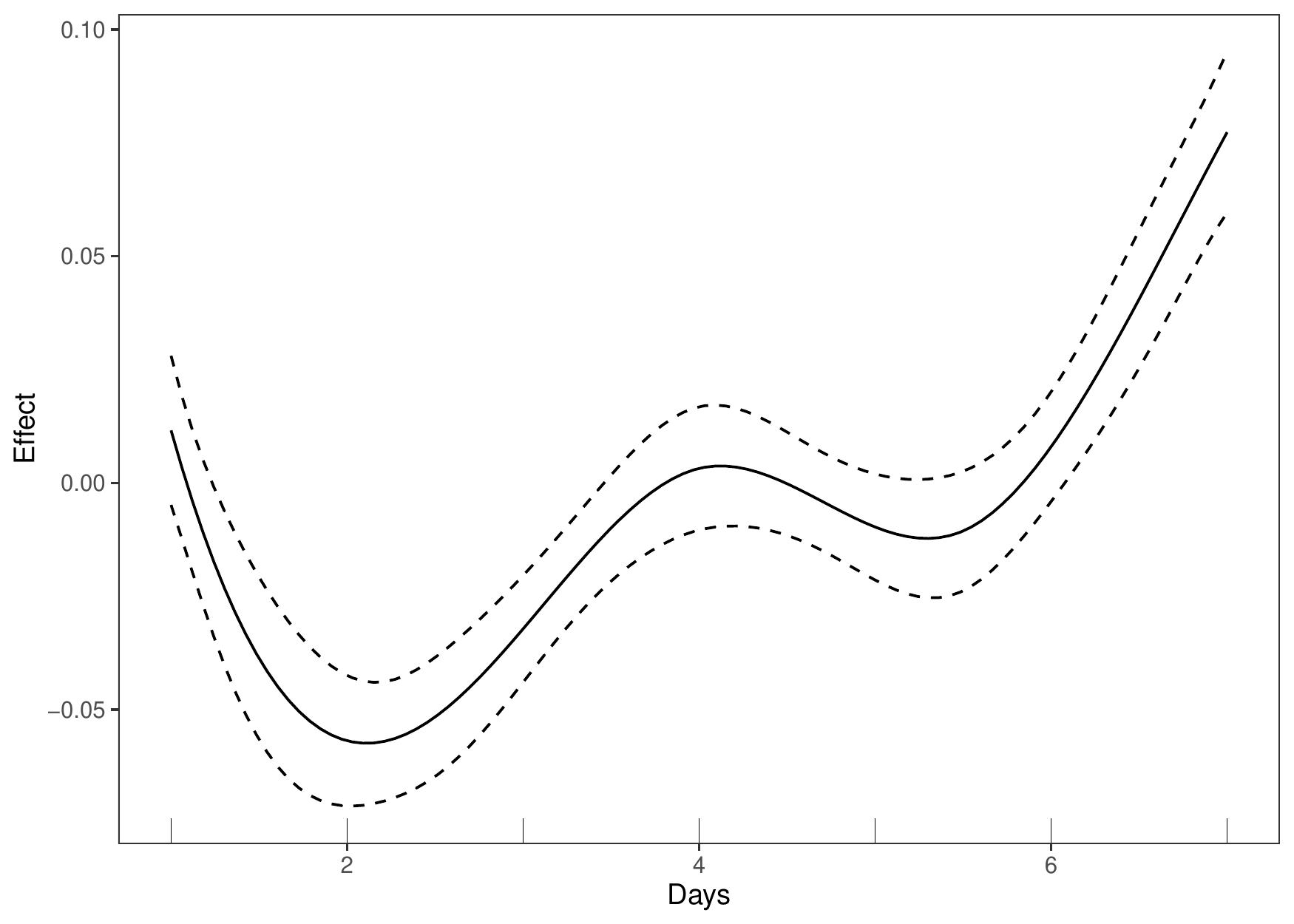}
		\caption{\label{fig11}}
		
	\end{subfigure}
	\begin{subfigure}[b]{0.4\textwidth}
		\includegraphics[width=\textwidth]{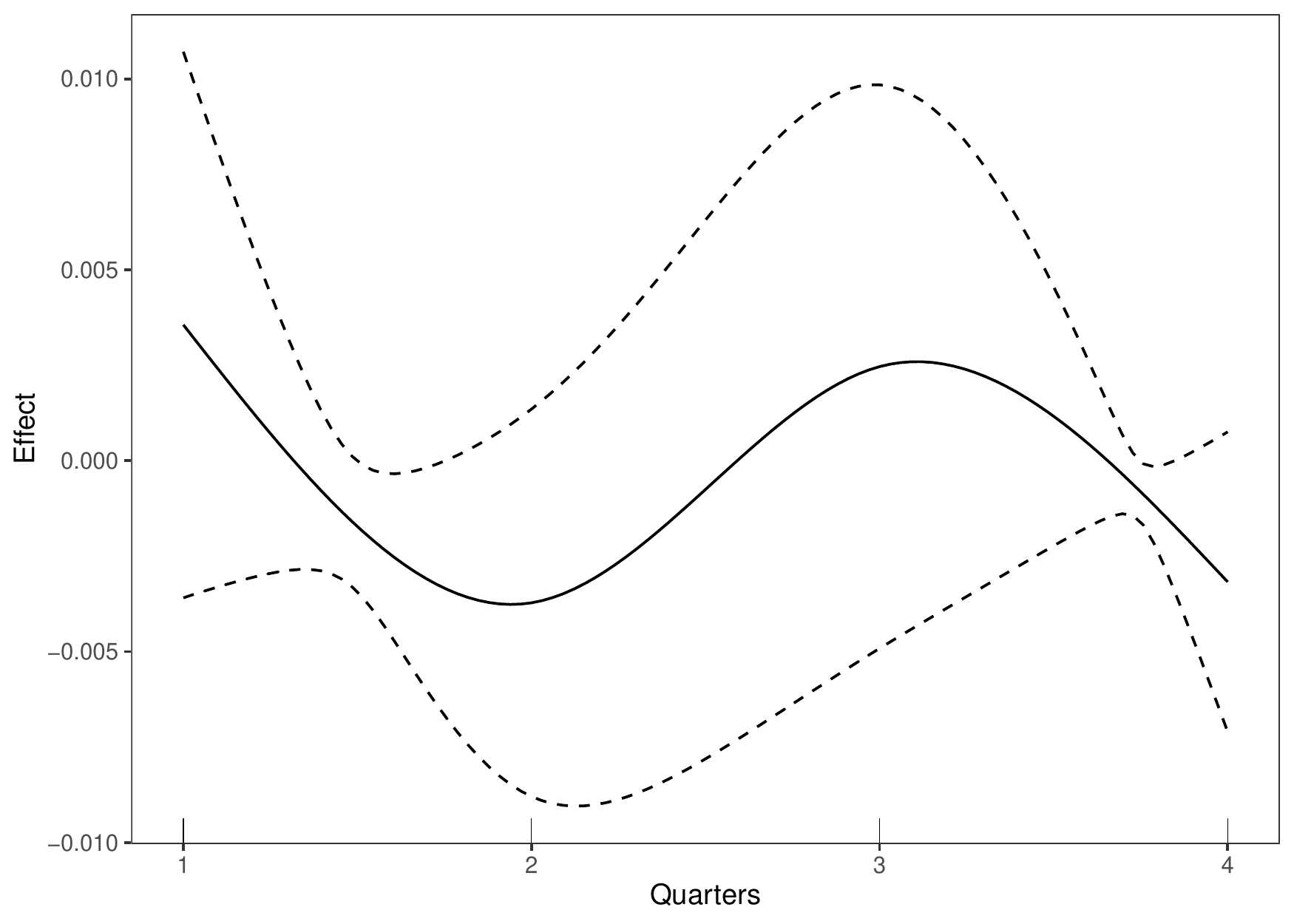}
		\caption{	\label{fig2}}
	
	\end{subfigure}
	\begin{subfigure}[b]{0.4\textwidth}
		\includegraphics[width=\textwidth]{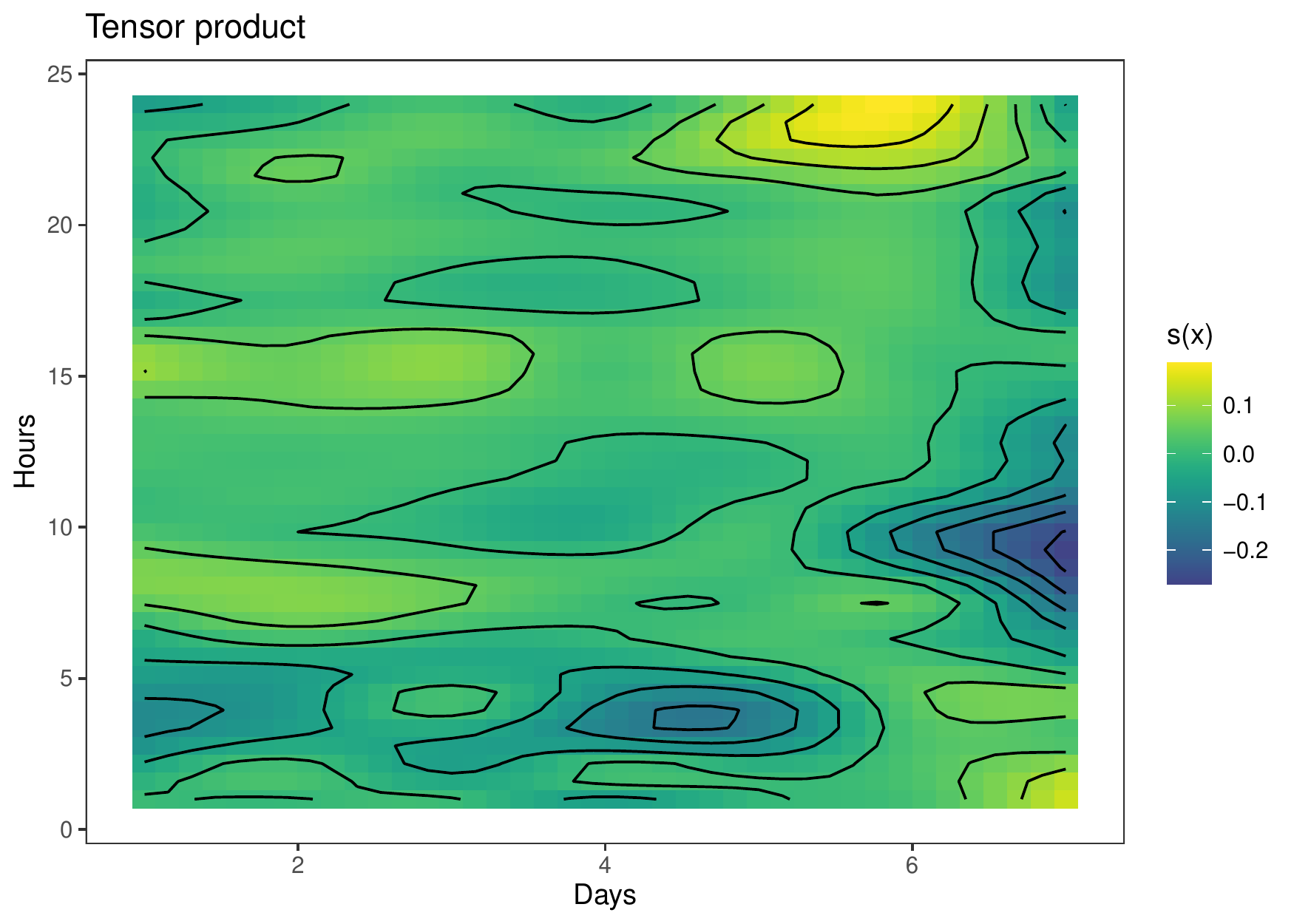}
		\caption{	\label{fig21}}
	
	\end{subfigure}
\caption{\emph{Alps} model coefficients plots during the COVID-19 pandemic: (a) Effect of the hour of the day; (b) Effect of the day of the week, where $1$ stands for Monday; (c) Effect of the quarters of the year, where $1$ stands for the first quarter of the year; (d) Effect of the interaction between days and hours, where $1$ stands for Monday.}
\label{fig:coef}

\end{figure}

\begin{figure}[H]
	\centering
	\begin{subfigure}[b]{0.4\textwidth}
		\includegraphics[width=\textwidth]{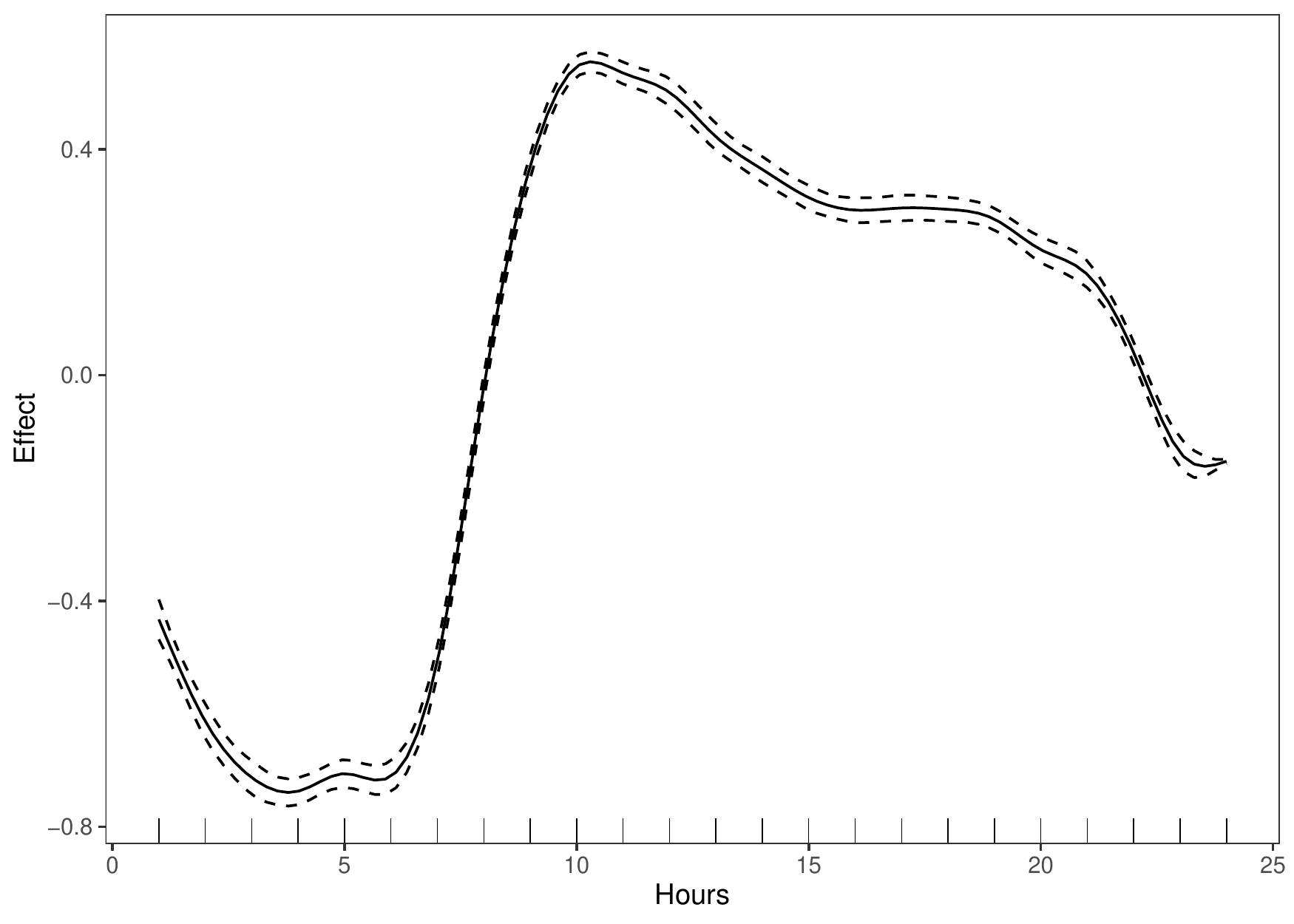}
		\caption{\label{fig1pre}}
		
	\end{subfigure}
	\begin{subfigure}[b]{0.4\textwidth}
		\includegraphics[width=\textwidth]{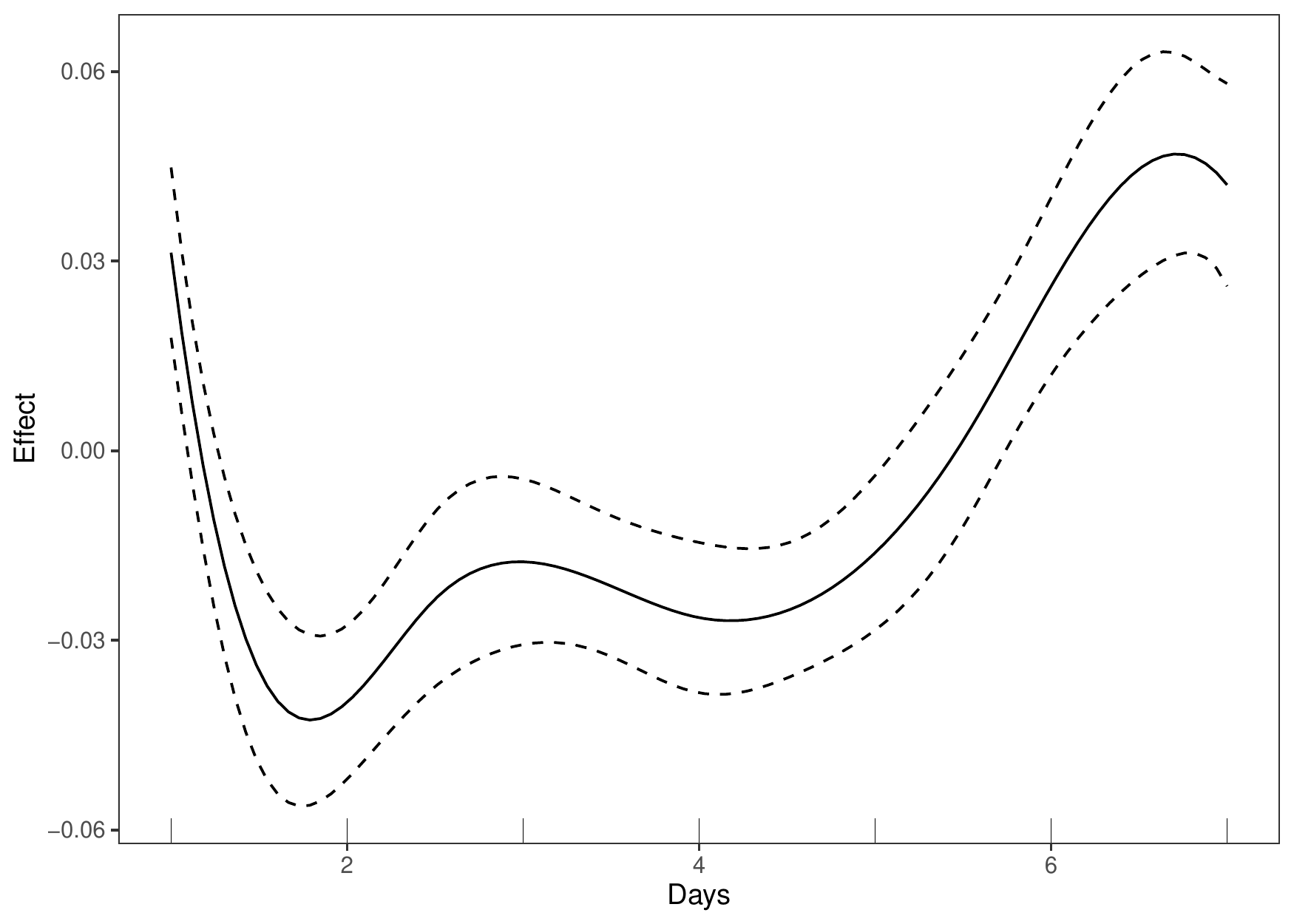}
		\caption{\label{fig11pre}}
		
	\end{subfigure}
	\begin{subfigure}[b]{0.4\textwidth}
		\includegraphics[width=\textwidth]{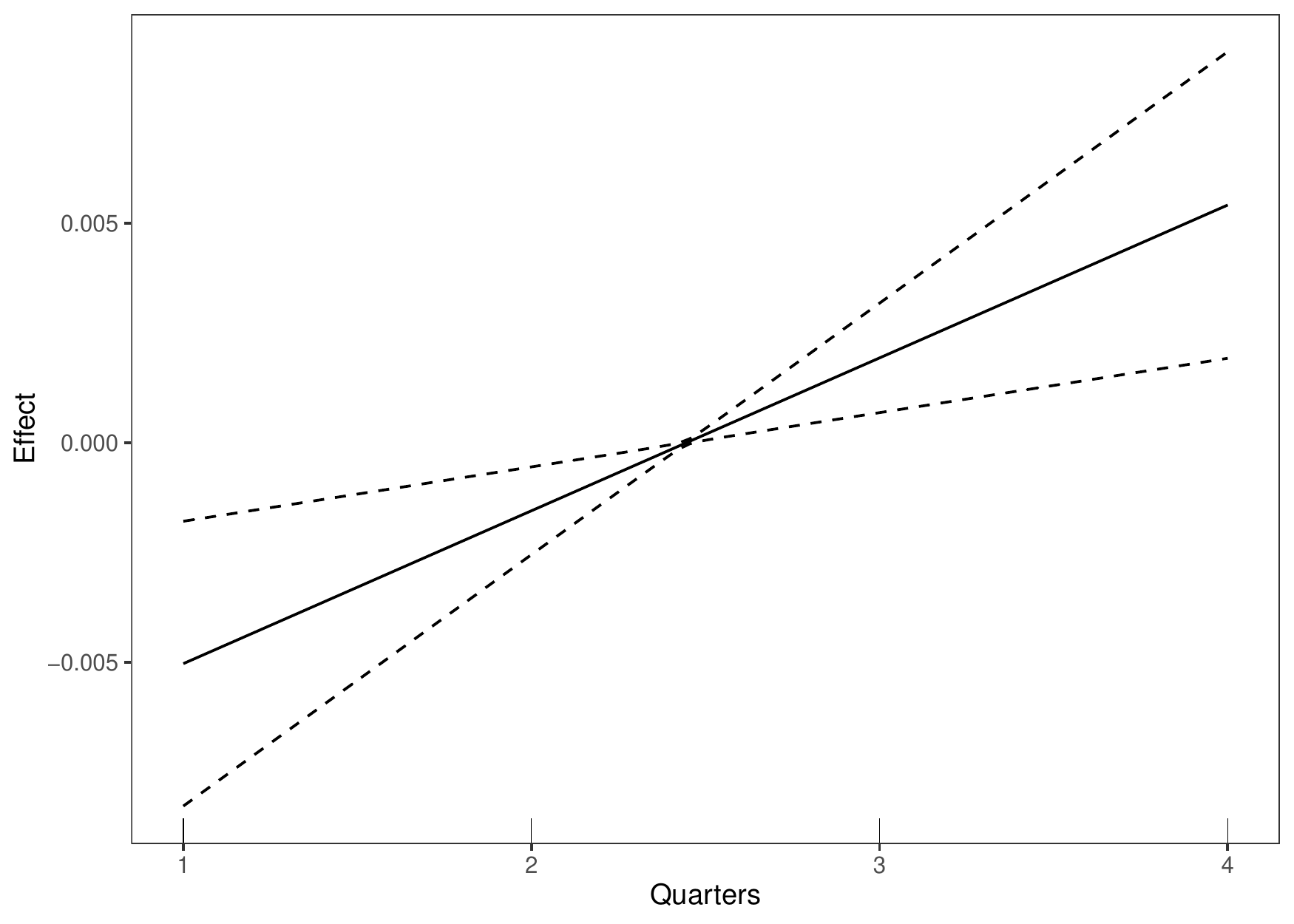}
		\caption{\label{fig2pre}}
		
	\end{subfigure}
	\begin{subfigure}[b]{0.4\textwidth}
		\includegraphics[width=\textwidth]{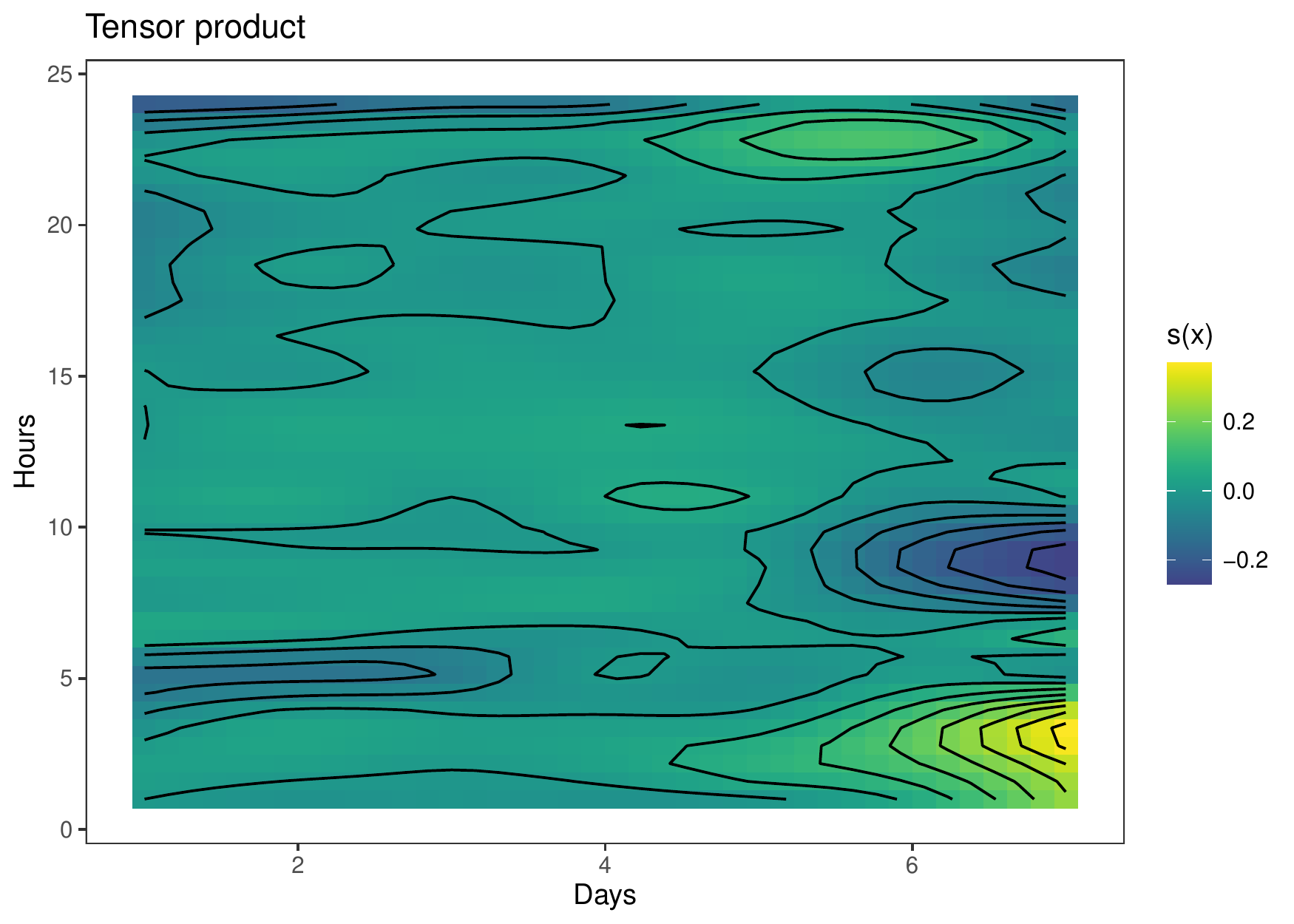}
		\caption{\label{fig21pre}}
		
	\end{subfigure}
\caption{\emph{Alps} model coefficients plots before the COVID-19 pandemic: (a) Effect of the hour of the day; (b) Effect of the day of the week, where $1$ stands for Monday; (c) Effect of the quarters of the year, where $1$ stands for the first quarter of the year; (d) Effect of the interaction between days and hours, where $1$ stands for Monday.}
\label{coef_preCovid}

\end{figure}


\end{document}